\documentclass[twocolumn]{pasj00}

\SetRunningHead{Astronomical Society of Japan}{Wide-field $^{12}$CO (1--0) imaging of M83
}

\Received{$\langle$2013 September 19$\rangle$}
\Accepted{$\langle$2013 December 25$\rangle$} 
\Published{$\langle$2014 May 1$\rangle$} 

\usepackage{ulem}

\begin{document} 
\title{
    Wide-field $^{12}$CO (J = 1--0) Imaging of the Nearby Barred Galaxy M83 with NMA and Nobeyema 45-m telescope: Molecular Gas Kinematics and Star Formation Along the Bar
}

\author{
Akihiko Hirota, \altaffilmark{1} \par
Nario. Kuno, \altaffilmark{1,2} 
Junichi. Baba, \altaffilmark{3}
Fumi. Egusa, \altaffilmark{4}
Asao. Habe, \altaffilmark{5}
Kazuyuki. Muraoka, \altaffilmark{6}
Ayako. Tanaka, \altaffilmark{7} 
Hiroyuki. Nakanishi, \altaffilmark{7}
Ryohei. Kawabe, \altaffilmark{8, 9, 2}
}
\altaffiltext{1}{Nobeyama Radio Observatory, Minamimaki, Minamisaku, Nagano 384-1805} \email{akihiko.hirota@nao.ac.jp}
\altaffiltext{2}{The Graduate University of Advanced Studies (SOKENDAI), 2-21-1 Osawa, Mitaka, Tokyo 181-8588} 
\altaffiltext{3}{Earth-Life Science Institute, Tokyo Institute of Technology, Ookayama, Meguro, Tokyo 152-8550}
\altaffiltext{4}{Institute of Space and Astronautical Science, Japan Aerospace Exploration Agency, Chuo-ku, Sagamihara, Kanagawa 252-5210}
\altaffiltext{5}{Department of Physics / Department of Cosmosciences, Hokkaido University, Kita-ku, Sapporo, Hokkaido 060-0810}
\altaffiltext{6}{Department of Physical Science, Osaka Prefecture University, 1-1 Gakuen-cho, Naka-ku, Sakai, Osaka 599-8531}
\altaffiltext{7}{Department of Physics, Graduate School of Science and Engineering, Kagoshima University, 1-21-35 Korimoto, Kagoshima, Kagoshima 890-0065}
\altaffiltext{8}{National Astronomical Observatory Japan (NAOJ), Osawa 2-21-1, Mitaka, Tokyo 181-8588}
\altaffiltext{9}{Department of Astronomy, The University of Tokyo, 7-3-1 Hongo, Bunkyo-ku, Tokyo 113-0033}

 \maketitle 
\begin {abstract} 
We present the results of the wide-field $^{12}$CO (1--0) observations of the nearby barred galaxy M83 
carried out with the Nobeyama Millimeter Array (NMA). 
The interferometric data are combined with the data obtained with the Nobeyama 45-m telescope to recover the total-flux.
The target fields of the observations cover the 
molecular bar and part of the spiral arms, with a spatial resolution of $\sim$110 pc $\times$ 260 pc. 
By exploiting the resolution and sensitivity to extended CO emission, 
the impact of the galactic structures on the molecular gas content is investigated 
in terms of the gas kinematics and the star formation.
By inspecting the gas kinematics, the pattern speed of the bar is estimated to be  
57.4 $\pm$ 2.8 km s$^{-1}$ kpc$^{-1}$, 
which places the corotation radius to be about 1.7 times the semi-major radius of the bar.
Within the observed field, HII regions brighter than 10$^{37.6}$ erg s$^{-1}$ in H$\alpha$ luminosity are 
found to be preferentially located downstream of the CO emitting regions.
Azimuthal angular offsets between molecular gas and star forming (SF) calculated with the angular cross-correlation method confirm the trend.
By comparing with a cloud orbit model based on the derived pattern speed, 
the angular offsets are found to be in accordance with a time delay of about 10 Myr.
Finally, to test whether the arm/bar promote star formation efficiency (SFE $\equiv$ Star Formation Rate (SFR)/H$_2$ mass), 
SFR is derived with the diffuse-background-subtracted H$\alpha$ and 24-$\mu$m images.
The arm-to-interarm ratio of the SFE is found 
to lie in the range of 2 to 5, while it is $\sim$1 if no background-removal is performed.
The CO-SF offsets and the enhancement of the SFE in the arm/bar found in the inner region of M83 are 
in agreement with the predictions of the classical galactic shock model.
\end {abstract}
\section{Introduction}
\label{sec: introduction}

The influence of galactic structures, namely spiral arms and galactic bars, 
on star formation and natal molecular clouds has long been a topic under active debate.
Discussions in this context often assume that 
galactic structures are quasi-stationary density waves (DWs) \citep{LinShu1964, BertinLin1996}, which are stationary over galactic rotations.
The response of gas clouds to stationary DW perturbations
has been investigated in detail, 
both for spiral arms (e.g., \cite{Fujimoto1968, Roberts1969}) 
and bars (e.g., \cite{Sorensen1976BarShock, Roberts1979BarShock}), 
and these models are collectively referred to as the "galactic shock" model.
\par
Within the picture of the stationary DW and the galactic shock models, gas clouds are continuously swept by the arm/bar, and these clouds subsequently form stars at an enhanced 
rate due to cloud-to-cloud collisions
(e.g., \cite{KwanValdes1983MassGrowth, Tomisaka1984ClCoagulation, Scoville1986ClCollision, RobertsStewart1987CloudCloudCollision}) 
and/or large-scale instabilities (e.g., \cite{BalbusCowie1985GravitationalInstability, Dobbs2008AgglomerationSelfGravity}).
These effects should produce two observable signs, 
which can be tested by molecular-line observations.
The first is the systematic offsets between molecular gas and star forming (SF) regions 
(e.g., \cite{VogelKulkarniScoville1988, Egusa2004, Foyle2011}).
The second is the enhancement of star formation efficiency (SFE),
which is defined as the star formation rate (SFR) per unit molecular gas mass, in the arm/bar compared to the interarm regions.
An alternative possibility has been also proposed for the case of bars (which is discussed later).
\par
Meanwhile, it has been recognized that at least some fraction of the spiral arms is not steady.
For example, a spiral arm which relies on swing amplification \citep{Toomre1981SwingAmp, SellwoodCarlberg1984, Baba2013} or self-propagating star formation \citep{MuellerArnett1976} should be non-steady in nature.
A non-steady arm behaves as material arm that winds up with time \citep{Wada2011Interplay, Grand2012BarSPH},  
and as a consequence of the co-moving of gas clouds and stars, 
produces neither ordered CO-SF offsets nor strong enhancement of SFE .
Thus, testing the two observable signs of the DW and the galactic shock models is of importance to interpret the nature of galactic structures.
\par
Thus far, observational verifications of the two abovementioned signs  
have been carried out mainly for spiral arms. 
For the CO-SF offsets, certain studies have indeed reported systematic offsets in grand design spiral galaxies 
(e.g., \cite{VogelKulkarniScoville1988, RandKulkarniRice1992, Egusa2004, Louie2013M51Offset}).
However, later, with an increase in the number of observed samples,
it is recognized that not all the galaxies show such offsets \citep{Egusa2009, Foyle2011}.
Compelling results as regards SFE enhancement also exist.
Although early works have reported elevated SFE in spiral arms 
\citep{CepaBeckman1990HaCoRatio, LordYoung1990M51, TacconiYoung1990, Knapen1996M10045m}, 
more recently, by comparing CO (2--1) and SFR traced 
using 24-$\mu$m and far-ultraviolet (FUV) images,
\citet{Foyle2010} showed that the arm-to-interarm variation 
in SFE is difficult to identify even in grand design spiral galaxies. 
\par
Compared to the number of tests performed for spiral arms, that for bars still remains small.
In a bar, the situation is complicated by the presence of highly non-circular motion.
Analyses of the CO-SF offsets in bars reported that to properly interpret the patterns of the offsets, 
detailed modeling of the gas kinematics is required \citep{Sheth2000NGC5383, Sheth2002, Koda2006NGC4303}, compared to the cases in spiral arms.
Moreover, it has been also argued that shear motion induced by the non-circular motion may stabilize clouds and lower the SFE, contrary to the case in spiral arms (e.g., \cite{Tubbs1982InhibitBarSF, Reynaud1998ShockInhibitedSF, Nimori2013M83}).
\par
Recent observational results concerning spiral arms that are not in agreement with the expectations of the classic DW and the galactic shock models 
\citep{Foyle2010, Foyle2011} 
might be attributed to the non-steady nature of spiral arms (e.g., \cite{DobbsPringle2010}). Within this context, bars are important targets for comparison with spiral arms, because they are in general considered to be long-lived compared to the arms, and thus, researchers can expect to observe the signs of the stationary DW, i.e. the CO-SF offsets and the SFE enhancement.

\begin {table} []
\begin {center}
\caption {Adopted parameters of M83}
\small
	\begin {tabular} {lr}
	\hline\hline
	Parameter& Value\\
	\hline
	Morph. \footnotemark[1]& SAB(s)c \\
	Center Position	(J2000)\footnotemark[2]& 13$^{\rm{h}}$ 37$^{\rm{m}}$ 00$^{\rm{s}}$.8\\
                                           & -29$^{\circ}$ 51$^{'}$ 56$^{''}$.0\\
	Position Angle$^{}$\footnotemark[3]& 225$^\circ$\\
	Inclination Angle$^{}$\footnotemark[3]& 24$^\circ$\\
	Systemic Velocity (LSR)\footnotemark[4]& 514 km\ s$^{-1}$\\
	Distance\footnotemark[5]& 4.5 Mpc\\
    Linear scale & 1$^{''}$ $\approx$ 22pc\\
    HI mass   \footnotemark[6]   &  7.7 $\times$ 10$^9$ M$_{\odot}$\\
    H$_2$ mass \footnotemark[7]  &  3.2 $\times$ 10$^9$ M$_{\odot}$\\
    E(B-V)\footnotemark[8]  & 0.070    \\
	\hline
	\end{tabular}
	\label {tab: m83_params}
\end {center}
{\small
	\footnotemark[1] {\citet{deVaucouleurs1991RC3}}\\
	\footnotemark[2] {IR center of \citet{SofueWakamatsu1994M83IRCenter}}\\
	\footnotemark[3] {\citet{Comte1981M83Parameters}}\\
	\footnotemark[4] {\citet{Kuno2007Atlas}}\\
	\footnotemark[5] {\citet{Thim2003M83Distance}}\\
    \footnotemark[6] {\citet{HuchtMeierBohnenstengel1981}, \\
    adjusted to the distance of 4.5 Mpc.}\\
    \footnotemark[7] {\citet{Crosthwaite2002M83}, \\
    adjusted to the distance of 4.5 Mpc.}\\
	\footnotemark[8] {\citet{Schlegel1998DustMap}}\\
}
\end {table}

\par
In this light, M83 is one of the most important nearby galaxies.
M83 is a nearby galaxy located at a distance of 4.5 Mpc \citep{Thim2003M83Distance} and 
this galaxy is one of the nearest face-on spiral galaxies that host prominent galactic structures, namely the bar and spiral arms.
Over the bar and arms, a pronounced pattern of dust lanes and HII regions (e.g., \cite{RumstayKauman1983}), which is also associated with a large number of young massive clusters (e.g., \cite{Larsen1999YMC, Chandar2010M83WFC3, Bastian2012M83}) and supernova remnants (\cite{Dopita2010, Blair2012}), is present.
\par
The central region of the galaxy hosts a bright starburst nucleus (e.g., \cite{Rieke1976Obs10um, Bohlin1983, TurnerHo1994VLA6cm2cm} ), and large scale stellar bar is likely responsible for feeding gaseous material to the nuclear region (\cite{Lundgren2004M83Kinematics, Fathi2008}), similar to the cases in other barred spirals (e.g., \cite{Ishizuki1990IC342, DownesReynaud1996NGC1530, Sakamoto1999, Regan1999, Sorai2000NGC253, Sheth2000NGC5383, Sheth2005BarInflow}).
Along the stellar bar, there are slightly curved, offset ridges which are 
associated with radio continuum argued to be related with shocked gas \citep{Ondrechen1985}.
\par
Since this galaxy has been known to be bright in CO emission, 
a large number of studies have been carried out at millimeter wavelengths 
with both single-dish telescopes 
\citep{Rickard1977CO, Combes1978M83, Wiklind1990M83, Handa1990, Crosthwaite2002M83, Lundgren2004M83Distribution, Muraoka2009M83ASTE} and interferometers \citep{LordKenney1991EasternArm, KenneyLord1991OrbitCrowding, RandLordHidgon1999M83, Sakamoto2004, Muraoka2009M83NMA}.
However, due to the galaxy's low declination, 
millimeter interferometric observations have had only limited $uv$ coverage with limited spatial sampling.
Thus, the detailed distribution of molecular gas in the inner galactic disk of M83 has not been fully clarified.\par
In this paper, 
we present results of a mosaic $^{12}$CO (1--0) observations of the nearby galaxy M83 
made via the Nobeyama Millimeter Array (NMA). 
To correct for the lack of the NMA's sensitivity to diffuse emission, 
the interferometric data are combined with the data obtained using the Nobeyama 45-m telescope.
The target of the observations cover the entire extent of the 
molecular bar and also part of the spiral arms with a spatial resolution
of $\sim$110 pc $\times$ 260 pc.
By exploiting the spatial resolution and ability to measure the total flux,
relations between molecular gas and star-forming regions are examined to verify whether galactic structures in the inner disk of M83 exert an influence on progress of star formation.  
Table \ref{tab: m83_params} summarizes parameters of M83 adopted throughout this paper.
\section{Observation and Data Reduction}
\begin {figure} []
  \begin {center}
  	\FigureFile(88mm, 88mm){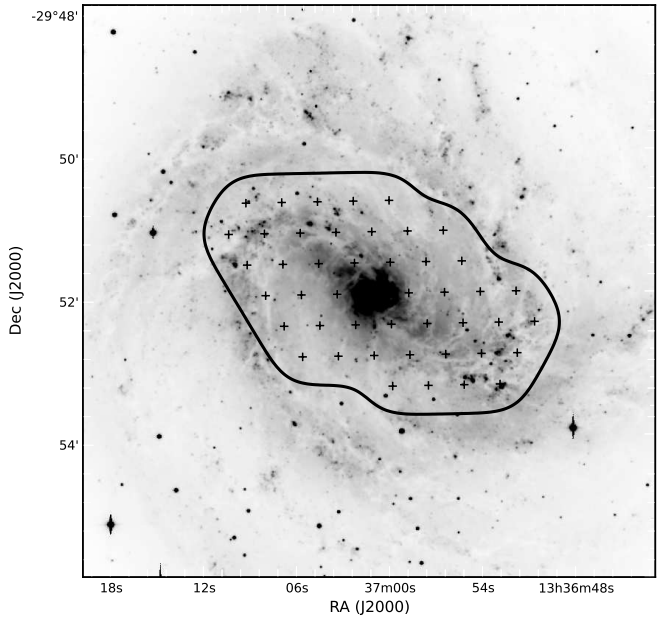}
	\caption {
        Optical $V$-band image of M83 retrieved from the NASA Extragalactic Database (NED) that was originally observed by \citet{Larsen1999YMC}.  The solid line is the contour of the normalized gain of the NMA mosaic observations at the level of 0.8.
    }
  \label {fig: obsregion}
  \end {center}
\end {figure}

\begin{table*}
    \begin{center}
    \caption{Properties of the NMA and the combined NMA+45-m data}
        \begin{tabular} {lrrrrr}
        \hline\hline
        Data set name& Beam size          & Beam position angle & r.m.s noise                         &                    & Total flux                               \\
        ..           & (arcsec)           & (degree)            &($\mathrm{mJy}\ \mathrm{beam}^{-1}$) & (mK) &               (K km s$^{-1}$)                          \\
        \hline
        NMA          & 11.1 $\times$ 4.7  & -8.8                & 57                                  & 101                & $\left(2.12 \pm 0.14\right) \times 10^5$ \\
        NMA+45m      & 12.0 $\times$ 5.5  & -10.7               & 66                                  &  92                & $\left(8.78 \pm 0.15\right) \times 10^5$ \\
        \hline
        \end{tabular}
    \label{tbl: data_param}
\end{center}
\end{table*}
\subsection{NMA Observation}
Aperture synthesis observations of the inner part of {M83} for $^{12}$CO(1--0) were 
carried out with the NMA. 
The NMA consists of six antennas with a diameter of 10 m, which provides a field-of-view size of $\sim$59$''$ at the rest frequency of $^{12}$CO (1--0).
During the observing periods, five antennas with the compact configuration (baseline length of 13-82 m) 
were operated.
S100 receivers, which operate in the double-sideband (DSB), were used as the front end and 
the UWBC spectrometer \citep{Okumura2000UWBC} that was 
configured to cover a 256-MHz bandwidth with 256 channels was used as the back end. 
Spectrometer outputs were subjected to the Hanning window function to suppress the Gibbs phenomenon.
The resultant frequency resolution was 2 MHz ($\sim$5.2 km s$^{-1}$).\par
Figure \ref{fig: obsregion} shows the target field of the NMA observations.
The target field covers the center, bar, and inner part of the spiral arms of the galaxy with 46 pointings separated with respect to each other by $\sim$30$^{''}$. 
The 46 pointings were observed sequentially with an integration time of 24s per pointing, and J1334-127
was observed once every 20 min as a gain calibrator.  
The sequential pattern of the pointings was changed several times to achieve uniform coverage of ($u,v$)-sampling points. 
The passband of the system was calibrated via the observation of bright quasars (3C273, 3C279) with an integration time of typically 45 min. 
Observations were carried out over 15 tracks in 2010 March and from January to March in 2011.
The system noise temperatures in the single-sideband (SSB) were typically 800-1300 K throughout the observations. \par
Raw visibility data were calibrated with the software package UVPROC-III (Kamazaki \& Morita, private communication), which is an IDL implementation of the UVPROC-II software package \citep{Tsutsumi1997UVPROC2}. 
Subsequent to the primary calibration processes including 
passband and complex gain calibrations, the visibility data were exported into the FITS format and further processing of the data, 
which includes short-spacing correction, imaging, deconvolution, and correction for primary-beam attenuation, were performed with the MIRIAD software package \citep{SaultTeubenWright1995MIRIAD}.
The resultant $^{12}$CO data cube has resolutions of 
11$^{''}$.2 $\times$ 4$^{''}$.73 and 
10.4 km s $^{-1}$ in the spatial and the velocity directions, respectively.
The rms noise within each channel was estimated to be $\sim$57 mJy beam$^{-1}$.
The total CO flux was calculated within 
an elliptical aperture with semi-major and semi-minor radii of 
130$^{''}$ and 75$^{''}$, respectively, which is 
inclined with a position angle of 145$^{\circ}$, and was determined as $\left(2.12 \pm 0.14\right) \times 10^5$ K km s$^{-1}$.
This is about 25 \% of the flux detected with the 45-m observations 
($\left(8.50 \pm 0.07\right) \times 10^5$ K km s$^{-1}$)
within the same aperture.
\subsection{NRO 45-m Telescope Observations}
\label{sec: obs_45m}
\begin {figure} []
  \begin {center}
  	\FigureFile(80mm, 80mm){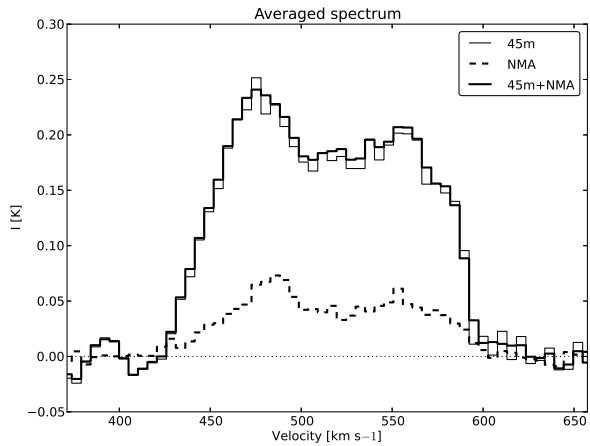}
	\caption {
        Global $^{12}$CO (1--0) profile averaged over an elliptical aperture with semi-major and semi-minor axes radii of 130$^{''}$ and 75$^{''}$, respectively, inclined with a position angle of 145$^{\circ}$. The thin solid line, heavy solid line and dashed lines indicate the averaged spectrum derived from the 45-m data, combined data, and NMA data alone, respectively.
    }
  \label {fig: averaged_sp}
  \end {center}
\end {figure}
As the NMA observations do not exhibit sensitivity to extended emission 
due to the central hole in array's ($u,v$)-coverage ($<4k {\lambda}$), 
the NMA data were combined with zero-spacing data obtained using the NRO 45-m telescope observations. 
As the front end, 25-BEam Array Receiver System (BEARS, \cite{Sunada2000}) which consists of a 5 $\times$ 5 focal-plane arrays of DSB, superconductor-insulator-superconductor (SIS) mixer receivers was utilized.
With the chopper-wheel method, 
intensity calibration in the DSB antenna temperature ($T_{\mathrm{A}}^*$) scale was performed.
Digital autocorrelation spectrometers \citep{Sorai2000SPIE} were used as back ends.
The spectrometers were configured to cover a 512MHz bandwidth with 1024 channels, and the Hanning window function was applied to the spectrometer outputs. \par
The observations were performed using the on-the-fly (OTF) observation mode that was implemented for the 45-m telescope by \citet{Sawada2008}.
The scanning rates were set between 40-50$^{''}$ s$^{-1}$, and data were sampled at a rate of 10 Hz, thereby providing a sufficient spatial sampling rate that met the Nyquist rate.  
To correct for the pointing offset, 
SiO J = 1-0 maser (42.821 and 43.122 GHz) sources were observed once every hour. 
Any data with pointing errors worse than 6$^{''}$ 
were excluded from the analysis. 
\par
The data reduction including flagging, applying scaling factors, and base-line subtraction was carried out using the NOSTAR software packages implemented at the NRO. 
A map was generated by convolving the observed data with a Gaussian-tapered Bessel function (see \cite{Sawada2008}) and 
regridding with a 8$^{''}$ grid along the spatial direction and a 2 MHz ($\sim$5.2 km s$^{-1}$) grid along the velocity direction. 
The resultant data cube had an effective resolution of $\sim$19$^{''}$.7 and an rms noise of $\sim$ 40 mK on the $T_{\mathrm{A}}^*$ scale. 
To convert the temperature scale from the $T_{\mathrm{A}}^*$ scale to the main-beam temperature, a DSB to SSB correction factor provided by the observatory and main beam efficiency ($\eta_{\rm{MB}}$) of 0.31 were applied. 
\subsection{Short Spacing Correction}
To combine the NMA data with the 45-m data, we converted the 45-m data cube into a set of pseudo-visibility data by following the method described by \citet{Kurono2009}. 
Throughout the combining process, the MIRIAD software package \citep{SaultTeubenWright1995MIRIAD} was utilized. 
 The overview of the combining process is as follows. 
 First, the 45-m data cube is deconvolved with a model beam that approximates the effective beam of the 45-m OTF observation.
Next, the deconvolved data cube is regridded to align with the center of each NMA pointing and subsequently, multiplied by a primary beam pattern of the NMA.
 Finally, to produce the visibility data, 
the Fourier transformation is applied to each data cube and sampled with sets of visibility data points that are randomly distributed on the $(u,v)$ plane within a range of 0 to 4.5 k$\lambda$. 
The pseudo-visibility data are combined with the NMA data and standard imaging procedures including inverse Fourier transformation, deconvolution with a dirty beam (CLEAN), and convolution of CLEAN components with a Gaussian beam are performed. 
\par
The resultant combined data cube had a spatial resolution of
12$^{''}$.0 $\times$ 5$^{''}$.54.
The rms noise within each channel was estimated to be $\sim$66 mJy beam$^{-1}$.
Figure \ref{fig: averaged_sp} shows comparison of the global CO (1--0) profiles of M83 made with the NMA-only, 45-m-only, and the combined data.
The total CO flux of the combined data was
$\left(8.78 \pm 0.15\right) \times 10^5$ K km s$^{-1}$ and this value is in agreement with that of the 45-m data.
Table \ref{tbl: data_param} summarizes the basic parameters of the data sets.\par
\citet{Crosthwaite2002M83} noted that peak value of the peak intensity at the nucleus was 890 mK for the 
55$^{''}$ beam. 
When the combined data were smoothed to the same 55$^{''}$ resolution, the peak temperature was 
884 $\pm$ 24 mK, and this is in good accordance with the abovementioned value.\par
Finally, two sets of the integrated intensity images were generated from the combined data.
The first image was a simple summation of the channel maps without any masking. 
The second was obtained by applying a mask that included any pixels with intensity above 4$\sigma$ and also included any pixels morphologically connected to the 4$\sigma$ kernels. 
The latter masked image was only used for visualization of the final image, which is made in \S\ref{sec: distribution} and \S\ref{sec: multi_comparison}.
For the remaining analyses that required intensity calculations (\S\ref{sec: angoffset} and \S\ref{sec: sfe}), the former unmasked image was utilized to avoid introducing statistical biases.

\section{Distribution and Kinematics of Molecular Gas}
\subsection{Distribution of $^{12}$CO (1--0)}
\label{sec: distribution}

\begin {figure} [ht]
    \FigureFile(88mm, 80mm){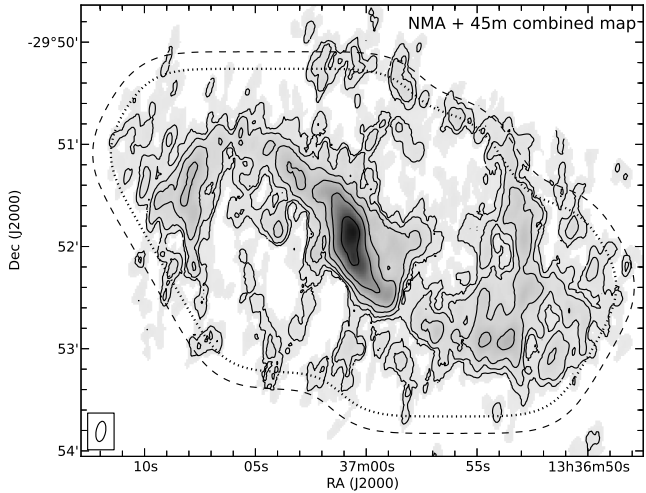}
    \caption {Velocity integrated map of the combined $^{12}$CO (1--0) data obtained from the NMA and the 45-m observational data.
    The contour levels are 8, 16, 32, 64, 128, and 256 K km s$^{-1}$, respectively. 
    Dashed and dotted curves indicate the normalized gain of the mosaic observations at the levels of 0.5 and 0.8, respectively.}
    \label{fig: m83_co_nma}
\end {figure}
\begin {figure} []
  \begin {center}
  	\FigureFile(88mm,80mm){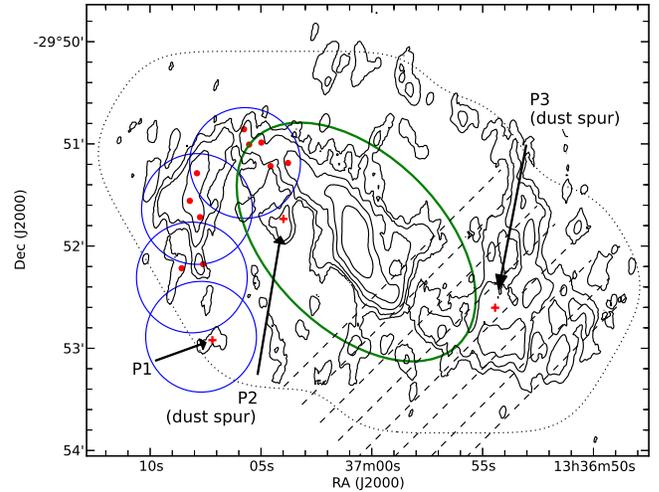}
	\caption {Integrated intensity map of the $^{12}$CO (1--0) emission overlaid with annotations to indicate the locations referred in the text. 
The dotted contour indicates the distribution of the mosaic gain to a unit point source, at the level of 0.4. 
Blue circles indicate the pointing centers of the OVRO observations made by R99 and red dots indicate the locations of the GMAs identified by R99. 
The green ellipse with a semi-major radius of 84$^{''}$ approximately indicates the extent of the stellar bar (see Appendix \S\ref{sec: hydro}). Dashed lines indicate the locations of the slits along which the position-velocity diagrams in figure 8 will be generated, later.
}
  \label {fig: guide}
  \end {center}
\end {figure}

Figure \ref{fig: m83_co_nma} shows the velocity-integrated intensity map of the combined $^{12}$CO (1--0) data.
The galactic structures including the bar and spiral arms are clearly resolved with a synthesized 12$^{''}$ $\times$ 5$^{''}$ beam ($\sim$260pc $\times$ 110 pc).
The molecular bar has a length of $\sim$200$^{''}$ ($\sim$4.4 kpc) and it lies at a position angle of $\sim$45 $^{\circ}$. 
Inside the bar, two ridges of molecular gas extend almost symmetrically with respect to the galactic center.
These ridges are located at the leading sides of the bar, and they gradually transit into spiral arms at both ends of the bar.
The transition zone between the bar and the spiral arm is located at the galactocentric radius range of 85$^{''}$ to 110$^{''}$. For the purpose of reference, the extent of the stellar bar is indicated in figure \ref{fig: guide} by an ellipse with a semi-major radius of 84$^{''}$.
\par
\begin {figure} []
  \begin {center}
  	\FigureFile(88mm,80mm){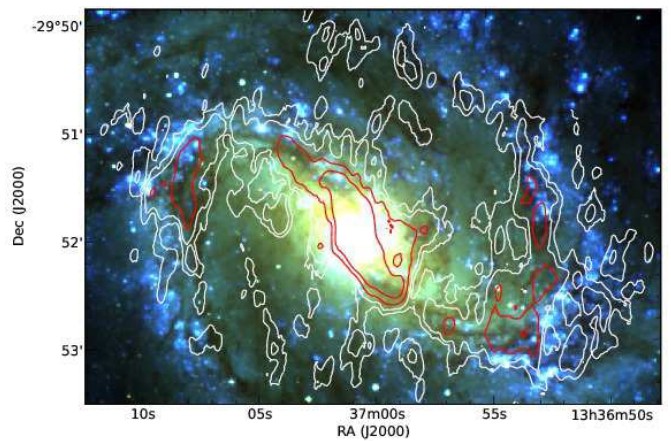}
	\caption {
$^{12}$CO integrated intensity image overlaid on a three-color composite image produced from the $I$-, $V$-, and $U$- band images \citep{Larsen1999YMC} assigned to red, green, and blue channels respectively. 
    }
  \label {fig: 3color_nma_co}
  \end {center}
\end {figure}

Inside of the bar radius, complex patterns of interarm clouds are observed. 
Figure \ref{fig: 3color_nma_co} shows a comparison of the CO image with the optical three-color composite image, made by assigning $I$-, $V$- and $U$-band images to red, green, and blue channels, respectively. The optical images were retrieved from the NASA Extragalactic Database (NED), and were originally released 
by \citet{Larsen1999YMC}. The distributions of the interarm clouds are well-correlated with the extinction features, and this indicates that interarm clouds seen in the NMA+45-m combined map are not artifacts. \par
Generally, interarm features in spiral galaxies are termed with different designations, depending on the locations of the features and the wavelength that is used to observe with.
In spiral arms, extinction features that extend from the arm toward the {\it leading} side is often called "feather" (see \citet{LaVigne2006} and references therein). 
On the other hand, barred galaxies show extinction features which extend from the main bar toward the {\it trailing} side, and they are termed as "dust spurs"  (e.g., \cite{Sheth2000NGC5383, Zurita2008}). 
The interarm features located at the trailing sides of the bar (P2 and P3 in figure \ref{fig: guide}) appear to correspond to the dust spurs. 
\par
Outside of the bar radius, two main spiral arms extend from both ends of the bar. The eastern end of the bar was previously mapped by \citet{RandLordHidgon1999M83} (hereafter R99) with the OVRO interferometer. 
For reference, the field of view of their observations and the location of giant molecular cloud associations (GMAs) identified by R99 are indicated in figure \ref{fig: guide}.
Despite the difference in beam sizes, the locations of the GMAs are in agreement with the combined CO image. 
R99 noted that while the CO distribution is generally well-associated with the dust lane near the end of the bar, 
the alignment deteriorates upon moving outward along the spiral arm.
In particular, they detected no CO emission near point 1 (P1) as indicated in figure \ref{fig: guide}, wherein the dust lane is visible (see figure \ref{fig: 3color_nma_co}).  Although several possible reasons for such misalignment were discussed by R99, it was concluded that none of them are solely acceptable.  Part of this ambiguity is due to the nature of the data; as they have also noted, their OVRO observations lacked the sensitivity to extended structures, and the noise level of the data ($\sim$0.1 K) was not sufficiently deep enough compared to the expected CO intensity estimated from the extinction value. In turn, although the noise level of the combined CO map is comparable to that of the R99 observations, the combined map recovers the sensitivity to diffuse emission.
The combined CO map shows the existence of the CO emission regions that are aligned with the dust extinction around P1, and thus, this map indicates the previous misalignment pointed out by R99 partly owing to the filtering nature of interferometric observations. 
\par
\subsection{Multi-wavelength Comparison}
\label{sec: multi_comparison}
\begin {figure*} [htbp]
    \FigureFile(180mm,160mm){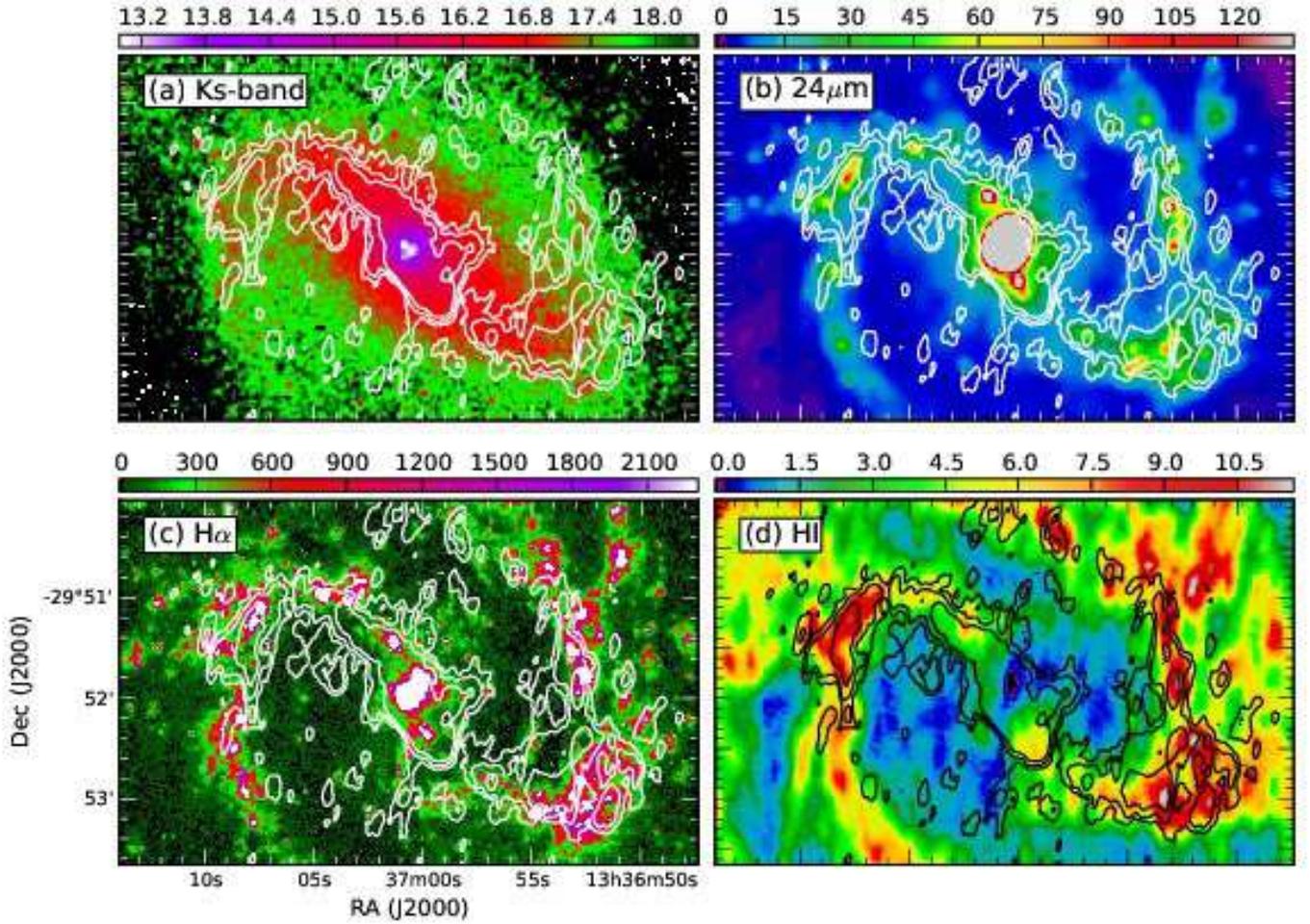}
    \caption {
        Comparison of the $^{12}$CO (1--0) image with multi-wavelength image sets.
        The CO image is indicated in each panel with contours.
        The levels of the CO contours are  10, 20, and 40 K km s$^{-1}$.
        (a) $K_s$-band (2.15-$\mu$m) image in units of magnitude.
        (b) 24-$\mu$m image in units of M Jy str$^{-1}$.
        (c) H$\alpha$ image in units of emission measure (EM). 
        (d) HI image in units of M$_{\odot}$ pc$^{-2}$. 
    }
    \label{fig: multic} 
\end {figure*}
In this subsection, the integrated intensity map of the combined $^{12}$CO (1--0) data is compared with images of the $K_s$-band, star formation tracers (H$\alpha$ and 24-$\mu$m), and atomic hydrogen
to study the relations between each tracer and molecular gas. \par
The $K_s$-band image was retrieved from the archive of the 2MASS Large Galaxy Atlas \citep{Jarrett2003}.
The H$\alpha$ image was retrieved from the data archive of the 
Survey for ionization in Neutral-Gas Galaxies
(SINGG; \cite{Meurer2006SINGG}).
The resolution of the image is limited by a seeing of $\sim$ 1$^{''}$.8. 
The 24-$\mu$m image taken with the MIPS camera \citep{Rieke2004MIPS} onboard the {\it Spitzer} {\it space} {\it telescope} \citep{Werner2004Spitzer} was retrieved from the {\it Spitzer} archive. 
The resolution of the 24-$\mu$m image is $\sim$5.7$^{''}$.
The HI data were obtained from the archive of The HI Nearby Galaxy Survey (THINGS; \cite{Walter2008THINGS}), which is a survey of HI carried out with the Very Large Array (VLA).
Images generated using the natural weighting of the visibilities were utilized, and synthesized beam with an elliptical Gaussian with FWHM sizes of $15^{''}.2$ $\times$ $11^{''}.4$, inclined with a position angle of 1$^{\circ}$.\par
Figure \ref{fig: multic}(a) shows the $K_s$-band image in comparison with the CO image.
Because of the small amount of extinction ($A_{K_s}$ $\sim$ 0.1$A_{\mathrm{V}}$), 
$K_s$-band emission is known to be an excellent tracer of 
underlying stellar mass. 
A stellar bar with a length of $\sim$84$^{''}$ (see Appendix \S\ref{sec: hydro}) is visible in the $K_s$-band image and molecular ridges inside the stellar bar are located at the leading sides.\par
Figures \ref{fig: multic}(b) and \ref{fig: multic}(c) show the distribution of CO (1--0) compared with star formation tracers.
Both images indicate that CO distribution at the 200-pc scale is well-correlated with star-forming regions.
The few exceptions are the interarm clouds located near the dust spurs where 
fewer signs of star formation are visible compared to the arm/bar regions.
Except for these interarm regions, the star-forming regions are consistent with the CO distribution, and in general, they seem to be preferentially located downstream of the CO arm/bar. 
If the gas clouds rotate faster than the underlying stellar potential, molecular material flowing from the trailing sides can be trapped and assembled at the arm/bar, and subsequently initiates star formation. 
The observed configurations appear to be largely in agreement with this scenario. 
To verify this scenario, 
the pattern speed of the bar is derived, and a more detailed analysis of the offsets will be made in \S\ref{sec: patternspeed} and \S\ref{sec: angoffset}.\par
Figure \ref{fig: multic}(d) shows the comparison of the HI image with CO (1--0).
Similar to the case of the star formation tracers, the distribution of HI is also well-correlated with CO (1--0). 
The surface mass density of HI is at most 11 M$_{\odot}$ pc$^{-2}$, and this value is well below that of CO. Thus, the inner galactic disk of M83 is predominantly molecular 
as pointed out by previous studies (\cite{Crosthwaite2002M83}, \cite{Lundgren2004M83Distribution}). 
Notably, the interarm CO clouds located in the trailing sides (dust spurs) are not bright in HI, and bright HI peaks are well-correlated with those of star-formation tracers. \par

\subsection{Molecular Gas Kinematics}
\label{sec: kinematics}
\begin {figure} []
    \FigureFile(80mm,80mm){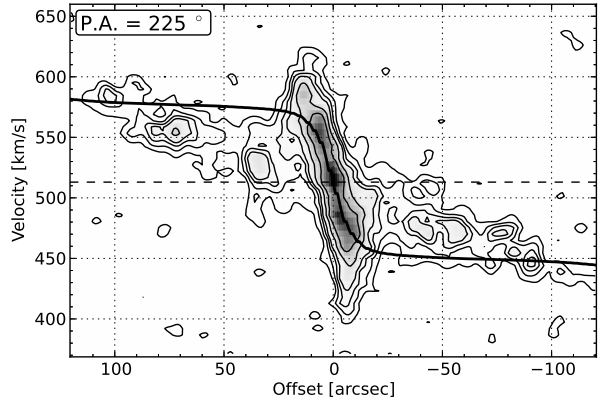}
    \caption {Position-velocity diagram along the major axis of M83 (PA = 225$^{\circ}$).  Contour levels are 2, 4, 6, 8, 12, and 16 $\times$ 110 mJy beam$^{-1}$. The solid curve indicates the model RC made by applying a constant mass to luminosity ratio to the $K_s$ distribution under the assumption of maximum disk contribution. The dashed horizontal line indicates systemic receding velocity of the galaxy ($V_{\rm{sys}} = 514$ km s$^{-1}$).
    }
    \label{fig: pvd_major} 
\end {figure}

Figure \ref{fig: pvd_major} shows the position-velocity diagram (PVD) along the major axis of M83.
In the PVD, multiple velocity components are present.
Within the galactocentric radius of 10$^{''}$, the variation in the line-of-sight (LOS) velocity indicates a steep rise in the rotation curve (RC), which reaches up to $\sim$300 km s$^{-1}$, when corrected for the inclination angle of 24$^{\circ}$. 
This steep rise in the nuclear RC was also reported by \citet{Handa1990}, and they speculated on the presence of a massive core.
Outside of the central region, the LOS velocity exhibits a fairly moderate gradient within the bar region.
The velocity width is fairly large within the bar (40-60 km s$^{-1}$, not corrected for inclination).
The overall trends observed in the PVD are in agreement with the previous CO observations
\citep{Handa1990, Crosthwaite2002M83, Lundgren2004M83Kinematics}. \par
Due to the close alignment between the position angles of the bar and the galaxy itself, 
bar-induced non-circular motion appears to severely affect the 
PVD along the major axis.
For reference, 
a model curve of the LOS velocity generated from the axisymmetric part of the stellar mass distribution is also indicated in figure \ref{fig: pvd_major}.
For the derivation of the stellar mass, see Appendix \S\ref{sec: hydro}.
Clearly, the observed LOS velocities are not in agreement with the circular rotational velocities, and consequently, this stresses the need for taking non-circular motions into account.\par
\begin {figure*} []
    \FigureFile(160mm,80mm){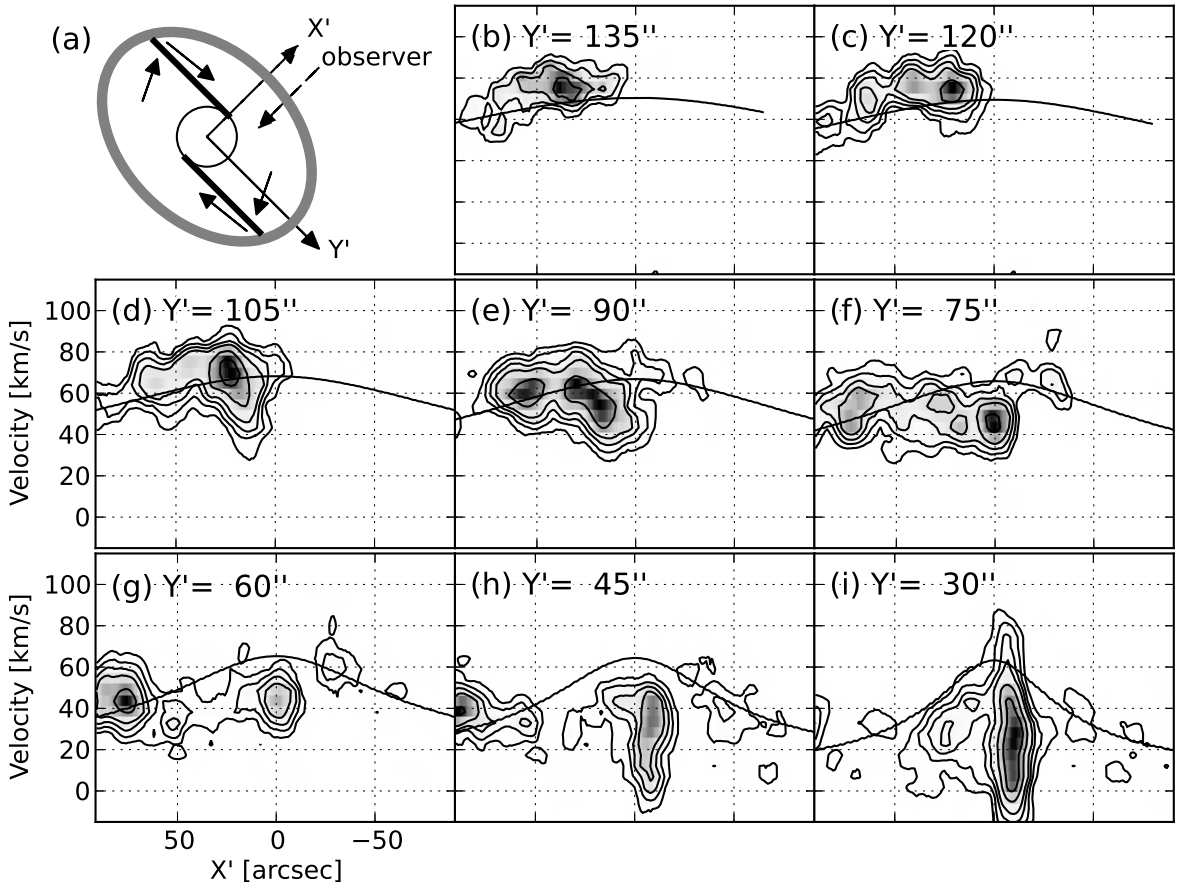}
    \caption {(a) Schematic diagram showing the configuration of the bar and 
    X'-Y' coordinates defined in the text. The ellipse and solid thick lines indicate the bar and offset ridges respectively. 
    Motion of gas clouds in the rotating frame with the bar pattern speed are also indicated with the arrows. 
    (b-i) Position-velocity diagrams perpendicular to the bar. The location of the slit  along which each PVD is generated, is also indicated in figure \ref{fig: guide}.
    Contour levels are 96mK $\times$ 2, 4, 6, 8, 12, and 16.
    The solid line in each figure indicates the model circular rotational velocity.
    }
    \label{fig: pvd_across_the_bar}
\end {figure*}

In general, the motion of gas clouds in a barred potential is thought to be 
 characterized by outward motion in the trailing sides that enters the offset ridges with a large open angle,
and inward motion that flows along the offset ridges in the leading sides (see figure \ref{fig: pvd_across_the_bar}a for reference). 
In the case of M83, tangential motion along the offset ridges minimally contribute to the LOS velocities as it is viewed side-on.
Thus, if PVDs are generated across the bar, we can expect to observe an 
abrupt change in the LOS velocity at the locations of the offset ridges.\par
Figures \ref{fig: pvd_across_the_bar}(b-i) show PVDs generated along the slits which runs perpendicular to the bar.
For convenience, here we define a rotated coordinated system ($X'$, $Y'$) in which $Y'$-axis is aligned to the position angle of the galaxy (225$^{\circ}$) (figure \ref{fig: pvd_across_the_bar}a). 
The PVDs in figure \ref{fig: pvd_across_the_bar} run parallel to the $X'$-axis, and they are separated from each other by 15$^{''}$. The locations of the slits are also indicated in figure \ref{fig: guide}.
Velocities are indicated as relative offsets to the systemic velocity of M83 (514 km s$^{-1}$). 
Trailing sides are located on the left side of the figures and thus gas clouds are expected to flow from the left to the right side.
The model circular rotational velocity which is the same as that for in figure \ref{fig: pvd_major} is also indicated as a reference.\par
Outside of the bar ($Y'$ = 135$^{''}$ and 120$^{''}$), the LOS velocities are largely in agreement with the circular velocities. 
As we move inward to the transitional zone between the bar and the spiral arm ($Y'$ = 105$^{''}$ and 90$^{''}$), 
the LOS velocities at the molecular ridge (near $X' = 0^{''}$) start to deviate from the circular velocity, 
thereby reflecting the fact that gas flows change their direction on the leading side.
Inside the bar ($Y'$ $\le$ $75^{''}$), the deviation from the rotational velocity becomes as large as 30-40 km s$^{-1}$.
The trends of the PVDs inside the bar and the bar-arm transitional zone are in agreement with the above expectation.

\subsection{Bar Pattern Speed}
\label{sec: patternspeed}
\begin {figure} []
  \begin {center}
  	\FigureFile(80mm, 90mm){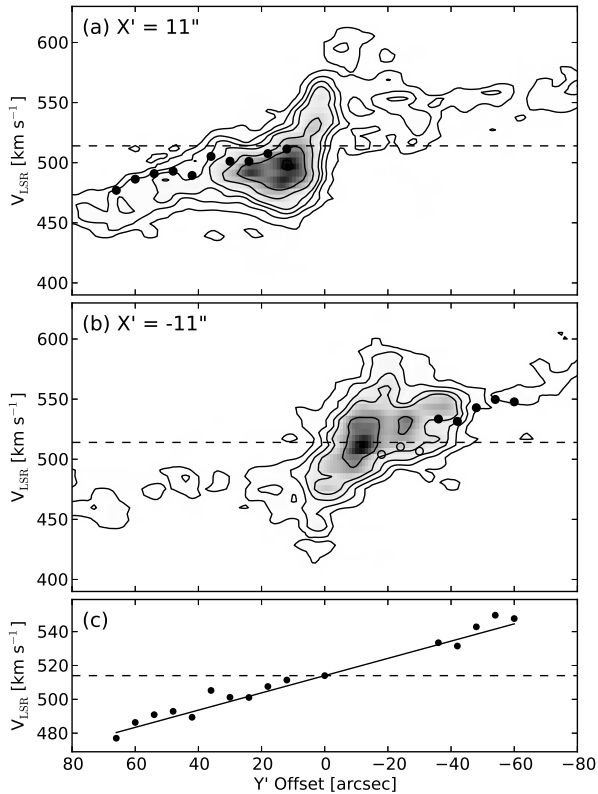}
	\caption {
        (a) PVD generated along $X'$ = -11$^{''}$.
        Contour levels are 1, 2, 3, \dots, and 6 $\times$ 1.2 Jy beam$^{-1}$. The dashed horizontal line indicates the 
        systemic velocity of M83 ($V_{\rm{sys}}$ = 514 km s$^{-1}$).
        Black dots indicate the lower envelope of the PVD, interpreted as the radial velocity of the clouds on the ridges.
        (b) same as (a), but for $X'$=11$^{''}$.
        White dots indicate the lower-envelope points, but they are not considered in the analysis 
        since located in the prohibited quadrant of the PVD.
        (c) Plot of the lower-envelope points adopted. The dashed line indicates the linear fit, and the slope yields the estimate of the pattern speed.
        }
  \label {fig: pvd_omegap}
  \end {center}
\end {figure}

From the previous subsection, we note that molecular gas kinematics within the bar of M83 are strongly affected by non-circular motions induced by the bar.
On the offset ridges, the observed LOS velocities are in agreement with the expectation that gas flows along the ridge. 
Exploiting this situation, it is possible to estimate the pattern speed of the bar \citep{Kuno2000}.\par
As the gas clouds located on the dust lane will move along it and 
as the dust lanes are seen side-on in the case of M83, 
the LOS velocity of molecular clouds along the offset ridges could be approximated as 
\begin {equation}
    V_{\rm{los}} \sim V_{\rm{sys}} + \Omega_{\rm{p}} \times R \cos\phi \times \sin{i}, 
\end {equation}
where $V_{\rm{sys}}$ denotes the systemic velocity, $R$ the galactocentric radius, $\Omega_{\rm{p}}$ the bar pattern speed, 
$\phi$ the angular offset with respect to the bar major axis, and $i$ the inclination angle of the galaxy
\citep{Kuno2000, Hirota2009}. 
For a PVD generated along an offset ridge, 
a rigid-rotation like feature is expected to appear since $R {\cos}{\phi}$ is equivalent to the distance offset along the slit. 
Although the accuracy of the measured pattern speed depends 
on the viewing angle and curvatures of the offset ridges,
this accuracy generally varies by $\sim$20\% \citep{Hirota2009}. 
Although the pattern speed of M83 had been estimated with this method in \citet{Hirota2009} using the CO data observed with the 45-m telescope \citep{Kuno2007Atlas}, here we will repeat the analysis with the combined NMA data that have finer resolution.
\par
Figures \ref{fig: pvd_omegap}(a) and \ref{fig: pvd_omegap}(b) show two sets of PVDs along the lines of $X'$ = 11$^{''}$ and $X'$ = -11$^{''}$ to trace the offset ridges. 
From each PVD, terminal velocities that define the lower envelope of the PVD are identified. To account for the effect of beam smearing, the terminal velocities at each position are defined as points where the intensity is 0.85 times the peak intensity at the position. 
Figure \ref{fig: pvd_omegap}(c) shows the plot of the terminal velocity as a function of spatial offset ($Y'$). 
A least-square fit was performed with the following linear expression
\begin {equation}
V = (Y' \times \Omega_{\mathrm{p}} - V_{\rm{sys}}) \sin{i}.
\end {equation}
The fit yields $\Omega_{\mathrm{p}}$ as 57.4 $\pm$ 2.8 km s$^{-1}$. \par
To verify the accuracy of the pattern speed derived here, 
a simple hydrodynamical simulation is performed.
In the simulation, fixed stellar potential estimated from the $K_s$ image is 
used and gaseous kinematics are solved with Eulerian hydrodynamical calculations. Details of the stellar mass derivation and hydrodynamical calculations are described in Appendix \ref{sec: hydro}.
The value of $\Omega_{\rm{p}}$ was chosen to be 57 km s$^{-1}$, which is close to the one determined here.\par
\begin {figure*} []
  \begin {center}
  	\FigureFile(160mm, 80mm){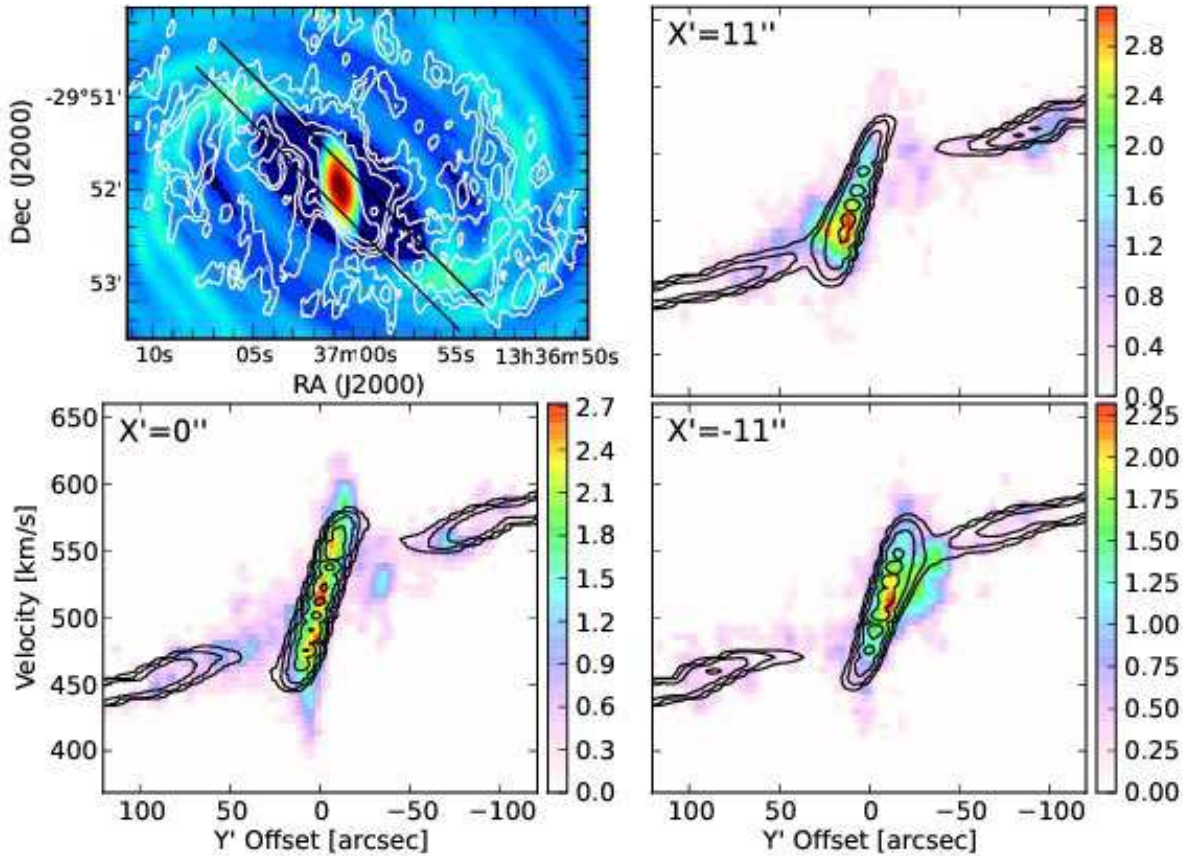}
	\caption{(a) Comparison of the observed and simulated data. Contour lines indicate the observed CO image, and the pseudo-colors indicates the simulated image. The two solid lines indicate the positions of the slits, $X'$ = 11$^{''}$ and $X'$ = -11$^{''}$. Simulated image is smoothed to the same spatial and velocity resolutions as those of the observed data. (b-d) Comparison of the PVDs generated from the observed (color) and simulated (contours) data, respectively. The two PVDS (b) and (d) are generated along the slits indicated in (a). The PVD in (c) is generated along the major axis of the galaxy.
  }
  \label {fig: model_obs}
  \end {center}
\end {figure*}

To compare with the observation,
the simulated hydrodynamical data
are assigned with the inclination and position angles of M83 (24$^\circ$ and 225$^\circ$, respectively) and then smoothed
to the same resolutions to the NMA+45m observations.
The Observed and simulated data are compared by
generating PVDs along the two offset ridges and along the major axis of the galaxy. Figure \ref{fig: model_obs} shows the comparison between the observed and simulated data.
The overall trends of the gas kinematics on the PVDs are well-reproduced by 
the model, and this supports the accuracy of the pattern speed determined here.\par
The pattern speed in this galaxy has been previously determined by several authors. 
\citet{LordKenney1991EasternArm} determined the location at which the H$\alpha$ arm crosses with dust lane,
and they assumed the point as the location of the corotation radius (CR). 
The pattern speed obtained by the author was $51$ km s$^{-1}$ kpc$^{-1}$, which corresponds to $\sim$57 km s$^{-1}$ kpc$^{-1}$ for the distance adopted here (4.5 Mpc).  
\citet{Zimmer2004} applied the TW method to $^{12}$CO (1--0) data obtained by 
\citet{Lundgren2004M83Distribution}, and they determined the pattern speed to be $45\pm8$ km\ s$^{-1}$\ kpc$^{-1}$, which corresponds to 
50 $\pm$ 9 km s$^{-1}$ kpc$^{-1}$ when corrected for the adopted distance.
The value of $\Omega_{\rm{p}}$ determined here is close to the values presented in previous works. 
Since the previous measurements measure the pattern speed of the spiral arm 
instead of the bar, the agreement of the measured values implies that both the bar and spiral arms share the similar angular rotational speed in M83.\par
\begin {figure} []
  \begin {center}
  	\FigureFile(88mm, 80mm){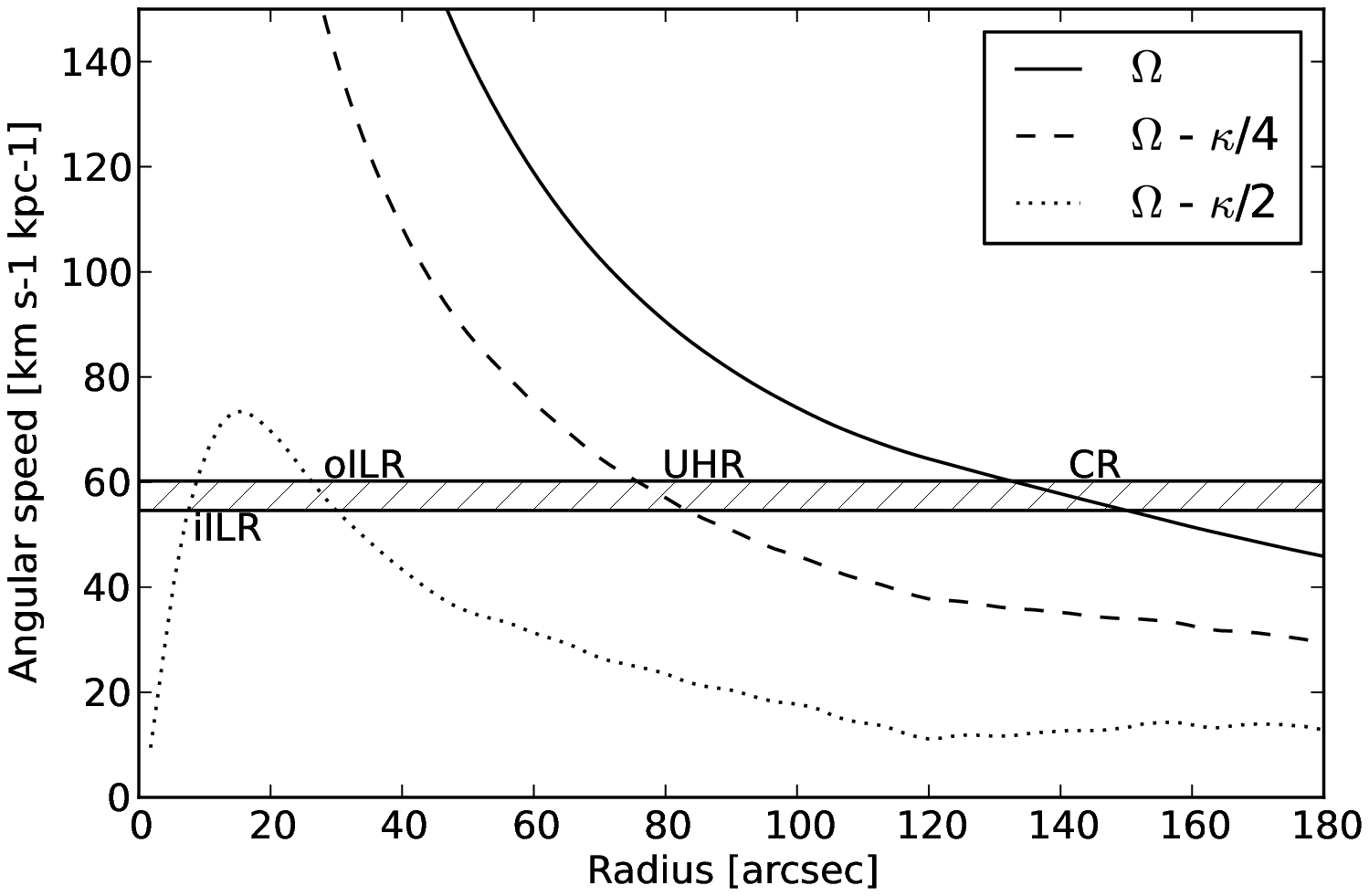}
	\caption{Angular velocities as a function of the galactocentric radius calculated from the circular rotational velocities estimated from the stellar mass distribution (Appendix \S\ref{sec: hydro}). Solid, dashed, and dotted lines indicate the value of $\Omega$, $\Omega-\kappa/4$, and $\Omega-\kappa/2$, respectively. The horizontal gray bar indicates the possible range of $\Omega_p$ values (57.4 $\pm$ 2.8 km s$^{-1}$ kpc$^{-1}$), determined here. The locations of the resonances, namely the inner Lindblad resonance (IILR), outer Lindblad resonance (OILR), and the ultra-harmonic resonance (UHR), and the corotation radius (CR) are indicated by labels.
    }
  \label {fig: ang_speed}
  \end {center}
\end {figure}

With the $\Omega_{\rm{p}}$ determined here and the circular rotational velocity estimated from the stellar mass distribution (Appendix \S\ref{sec: hydro}), 
the locations of the orbital resonances are determined (figure \ref{fig: ang_speed}).
The CR is located at the galactocentric radius of 141 $\pm$ 14$^{''}$ and is far away from the end of the bar ($\sim$85$^{''}$).
This is in contrast to the traditional assumption that the CR is located at the end of the bar.
On the other hand, the ends of the bar are fairly close to the inner 4:1 (UHR) resonance, which is located at the galactocentric radius of 80 $\pm$ 6$^{''}$.

\section{Derivation of Star Formation Rate}
\label{sec: sfr}
In this section, we derive star formation rate (SFR) in M83 using the H$\alpha$ and 24-$\mu$m images presented in 
\S\ref{sec: multi_comparison}. To correct for contamination by the diffuse background emission that are not related to massive star formation, we morphologically separate the discrete emission associated with star-forming regions from the underlying diffuse emission. 
\par

In deriving the SFR for unresolved extragalactic studies, 
corrections for extinction and subtraction of smooth background emission not related to star formation are essential. 
As the extinction diminishes the H$\alpha$ flux by a factor of 3 to 10 or more with typical extragalactic observations
(e.g., \cite{Scoville2001M51, Prescott2007}), 
considerable attention has been focused on extinction correction.
One of the most popular approaches utilized nowadays is to combine the H$\alpha$ image with the corresponding 24-$\mu$m image obtained using the MIPS camera onboard the {\it Spitzer space telescope} \citep{Werner2004Spitzer} to trace both unobscured and obscured high-mass star formation.
This approach is particularly attractive in that it simply provides an empirical relation that relates the SFR to the linear combination of the H$\alpha$ and 24-$\mu$m fluxes \citep{Calzetti2007, Kennicutt2007ApjM51}. \par

\par
However, the use of the 24-$\mu$m emission as a tracer of SFR likely complicates the problem of background removal, 
and few studies have paid attention to this aspect.
Both H$\alpha$ and 24-$\mu$m flux from galaxies contain a fraction of diffuse emission not directly related to recent star formation.
Diffuse H$\alpha$ emission, often referred to as diffuse ionized gas (DIG) emission, is considered to be mainly originating from the escaped Lyman continuum of the HII regions (e.g., \cite{Ferguson1996}), diffusing over the galactic disk.
A fraction of such diffuse H$\alpha$ is found to be 30-50 \% of the total H$\alpha$ flux (e.g., \cite{Thilker2000}; \cite{Oey2007DIG}) although \citet{Crocker2013} also reported that fraction reduces if differential-extinction is taken into account.\par
Meanwhile, diffuse 24-$\mu$m emission is considered to be 
originating from 
small dust grains not in temperature equilibrium and stochastically powered by the interstellar radiation field through single-photon heating
(e.g., \cite{Desert1990, DraineLi2007}). 
As opposed to the case of the H$\alpha$, 
fraction of the diffuse emission contamination for the 24-$\mu$m band has not been studied in detail, but is speculated to be larger compared to the fraction of diffuse H$\alpha$ emission (e.g., \cite{Liu2011KS}).
\par
Originally, the calibration of the 24-$\mu$m as an SFR tracer was made via aperture photometry of extragalactic HII regions (\cite{Calzetti2005ApjM51, Calzetti2007, Kennicutt2007ApjM51}), in which diffuse emission was removed as local background.
However, subsequent pixel-based studies utilizing the 24-$\mu$m calibration often 
have not considered the factor of diffuse emission.
Only a few recent studies have addressed the removal of diffuse background emission (\cite{Rahman2011NGC4254, Liu2011KS, Leroy2012, Momose2013}).
\citet{Rahman2011NGC4254} removed the slowly varying diffuse emission by utilizing unsharp masking, and \citet{Liu2011KS} and \citet{Momose2013} separated discrete emission arising from the HII regions from the background by using the source extraction method.
Besides these morphological approaches, \citet{Leroy2012} incorporated spectral information along with the dust model.
Here, we adopt the method utilized by \citet{Liu2011KS}. \par
In \S\ref{sec: remove_de_halpha} and \S\ref{sec: remove_de_24um}, discrete sources regarded as HII regions (or associations of HII regions) are identified from the H$\alpha$ and 24-$\mu$m images, respectively.  Using the background-removed images, we derive the SFR in \S\ref{sec: calc_sfr}, and the radial distribution of SFR and SFE is presented in \S\ref{sec: radial}.

\subsection{Removal of Diffuse Emission from H$\alpha$ Image}
\label{sec: remove_de_halpha}
\begin {figure} []
    \FigureFile(88mm,90mm){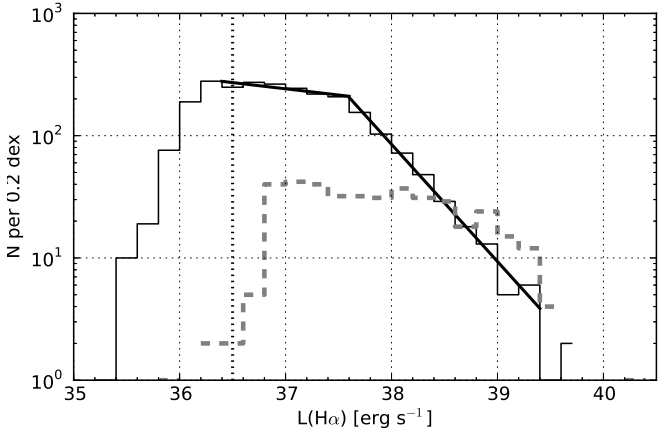}
    \caption {H$\alpha$ luminosity function (LF) for the identified HII regions in M83. H$\alpha$ luminosity has not been corrected for internal extinction (solid line).
    The dotted vertical line at 10$^{36.5}$ erg s$^{-1}$ indicates the estimated completeness limit of the HIIphot run.
    The two solid lines indicate the fit to the HII LF.
    In addition, the dashed line indicates the LF derived from the smoothed H$\alpha$ image with a resolution of 6$^{''}$.
    }
    \label{fig: hii_lf}
\end {figure}
\begin {figure*} []
    \FigureFile(160mm,90mm){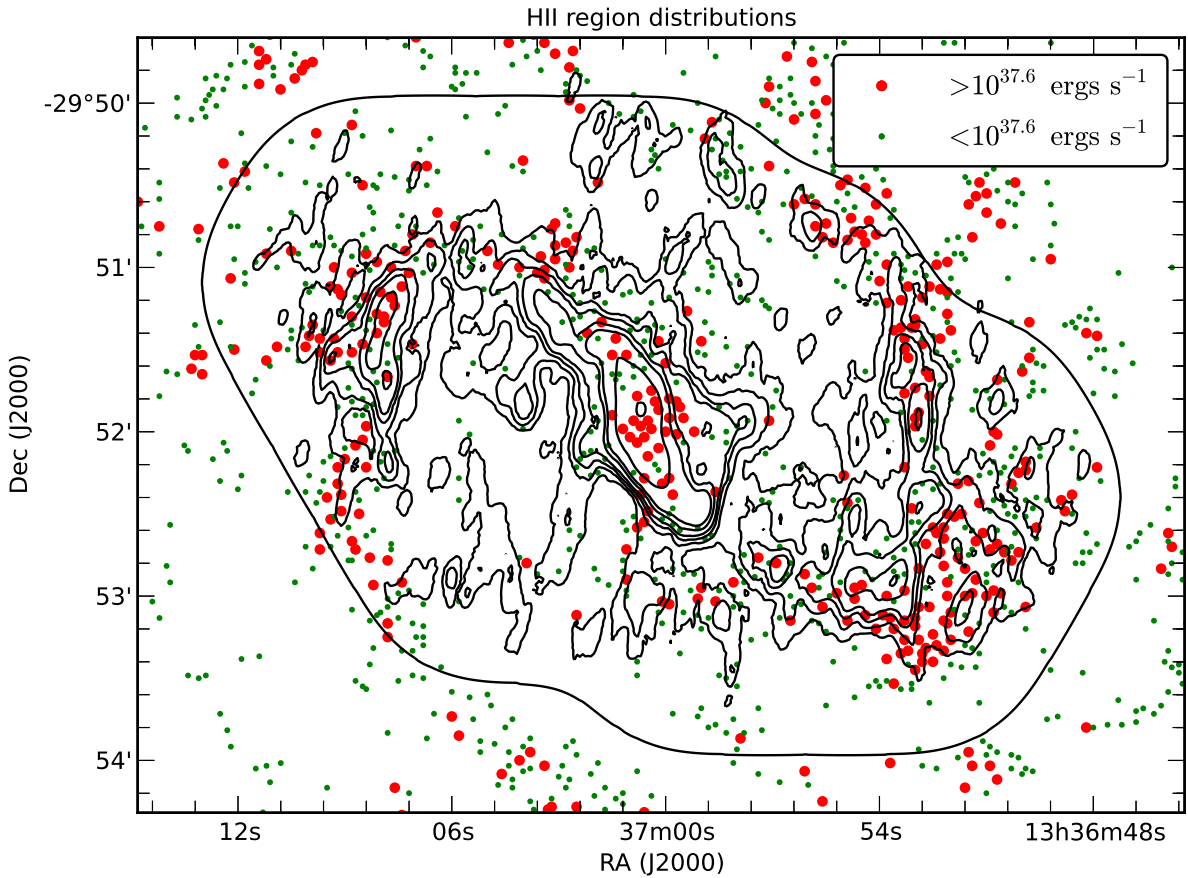}
    \caption {Locations of the HII region candidates identified with the HIIphot code (\cite{Thilker2000}). 
    HII regions with luminosity (uncorrected for internal absorption) larger than 10$^{37.6}$ erg s$^{-1}$ are indicated by red circles and those below the threshold are identified by green dots.  Contour levels are 10, 20, 30, 40, 60, 80, 160, and 320 K km s$^{-1}$.
    }
    \label{fig: m83_hiiregions}
\end {figure*}

To isolate the discrete emission arising from the HII regions, we utilized the HIIphot code \citep{Thilker2000}. 
As the full description of the algorithm is available in \citet{Thilker2000}, we only present the outline here.
The HIIphot algorithm works in the following manner.
First, it identifies local maxima from the image as HII-region candidates.
To account for sources with various sizes, 
the source identification procedure is performed for several sets of images that are generated by convolving the original image with various sizes of the convolution kernel.
Since the candidate sources identified in this step only trace the emission peaks of HII regions, they are treated as "seed" regions.
Each such "seed" candidate is fitted with models of HII-region distributions that exhibit Gaussian or ring morphologies, and the 
``the goodness of fit'' is evaluated by calculating a modified version of the Pearson's linear correlation coefficient ($\rho$).
Next, the "seed" candidates are sorted in descending order of $\rho$, and only the sources with the highest $\rho$ value at each position are considered.
Subsequently, each seed region is expanded until the intensity gradient reaches a user-specified threshold value, which is referred to as the terminal gradient.
After the boundaries of the HII regions are determined, 
the code produces a background image by performing two-dimensional fitting to the boundaries and then 
replacing the regions inside the boundaries with the result of the fitting.
Using the determined boundaries and the background image, 
photometry for each HII region is finally performed.  
\par
The H$\alpha$ and associated continuum images from the SINGG survey \citep{Meurer2006SINGG} were utilized as inputs to the HIIphot code \citep{Thilker2000}.
Correction for foreground extinction was made using the the value ($A_{\mathrm{H}\alpha}$ = 0.18) presented by \citet{Meurer2006SINGG}.
The contamination of the [NII] lines at rest wavelengths of 6548 \AA \ and 6583 \AA \ were also made by the calculated fractional contribution of [NII] lines to the total bandpass (0.11, \cite{Meurer2006SINGG}). The application of these corrections resulted in the scaling the the original image by a factor of $\sim$1.05. \par
The detection limit of the local maxima was set to 5$\sigma$ and the terminal gradient was chosen to be 6 EM pc$^{-6}$.
Although the use of the value of 1.5 EM pc$^{-6}$ was recommended by \citet{Thilker2000}, 
by examining the final extracted sources, we estimated that the value of 6 EM pc$^{-6}$ is appropriate for the data utilized here. 
We also note that \citet{Oey2007DIG} have also applied the HIIphot code to the same data set and they adopted the same terminal gradient. 
\par
Figure \ref{fig: hii_lf} shows the observed (not corrected for internal extinction) luminosity function (LF) of the identified HII regions. 
The completeness limit of our HIIphot run was estimated by 
randomly distributing artificial sources with sizes between 5 to 30pc in the uncrowded portion of the image 
and performing HIIphot extraction. 
By examining the detection rate of the input sources, 
we estimated the completeness limit to be 36.5 erg s$^{-1}$.\par
To characterize the LF, it is fitted with the following power-law function:
\begin {equation}
    {d{\cal N}} = A L^{\alpha} {d{\ln}L}, 
\end {equation}
where ${\cal N}$ and $L$ denotes number and luminosity of HII regions, respectively, and $A$ represents a constant. 
There is a break in the LF curve at 37.6 erg s$^{-1}$, and the index ($\alpha$) is -0.93 $\pm$ 0.13 and -0.10 $\pm$ 0.08 above and below the threshold, respectively.
If we set $A_{\mathrm{H}\alpha}=1.4$ \citep{Prescott2007} as the nominal internal extinction, 
the luminosity of 10$^{37.6}$ erg s$^{-1}$ approximately corresponds to the suggested maximum luminosity limit of a single-star HII region (about 10$^{38}$ erg s$^{-1}$, e.g., \cite{KennicuttHodge1986LMC, Scoville2001M51, Lee2011M51}). 
Thus, we interpret that HII regions brighter than 10$^{37.6}$ erg s$^{-1}$ are powered by 
massive clusters that includes several tens of O and B stars of O and B stars, 
rather than low-mass clusters that is powered by several OB stars.
\par
Figure \ref{fig: m83_hiiregions} shows the distribution of the identified HII regions. 
While the bright HII regions ($\ge$10$^{37.6}$ erg s$^{-1}$) are strongly concentrated around the bar and the arms, 
the dimmer ones are distributed almost uniformly. 
Notably, bright samples are also preferentially located downstream of the CO peaks. 
This point is discussed in \S\ref{sec: angoffset}. \par
The HIIphot code produces a background image that is generated by subtracting the identified sources and filling in the residual holes with surface fitting.
The total flux of the background image within the circular aperture with a radius of 430$^{''}$ centered on M83 is $\sim$4.3 $\times$ 10$^{40}$ erg s$^{-1}$. 
The 1$\sigma$ uncertainty considering only the rms noise is 2.3 $\times$ 10$^{38}$ erg s$^{-1}$. 
The fractional amount of background with respect to the total emission is 0.25, and this value is comparable to the value of $\sim$ 0.4 found in other nearby galaxies (e.g., \cite{Thilker2000, Oey2007DIG}).\par
Since the 
The background image should still contain a fraction of low-luminosity HII regions under the completeness limit of 10$^{36.5}$ erg s$^{-1}$.
We estimated the fraction of undetected low-luminosity HII regions by making the following assumptions: 1) the LF has a fixed slope under 10$^{37.6}$ erg s$^{-1}$, 
and 2) the minimum luminosity of the HII region is 10$^{34}$ erg s$^{-1}$, 
which corresponds to the luminosity of HII regions powered by of a single B0-B1 star (e.g., \cite{Azimlu2011M31}).
By integrating the expected and the observed LF within the range between 10$^{35}$ and 10$^{36.5}$ erg s$^{-1}$, 
we find that total amount of the missing fraction of the 
undetected HII-region flux amounts to 7.9 $\times$ 10$^{38}$ erg s$^{-1}$. 
As this value is well below the summation of the background flux, we conclude that nearly all
the HII regions contributing to the production of ionizing photons have been accounted in our calculations.\par
Finally, by subtracting the background image from the original H$\alpha$ image and masking certain regions around the foreground stars, we generated a background-subtracted H$\alpha$ image.
Figures \ref{fig: remove_de}a and \ref{fig: remove_de}b show the original and background removed H$\alpha$ images respectively, with the same color scale.

\subsection{Removal of Diffuse Emission from 24-$\mu$m Image}
\label{sec: remove_de_24um}
To isolate the discrete emission features from the 24-$\mu$m image, 
the H$\alpha$ image smoothed to the same resolution as the 24-$\mu$m image was utilized as a reference to define the boundaries of the HII regions. 
The HIIphot extraction was performed on the smoothed H$\alpha$ image with an adopted terminal gradient of 1 EM (dashed line in figure \ref{fig: hii_lf}).
The background H$\alpha$ image generated in this step exhibited a total flux of $\sim$4.7$\times$10$^{40}$ erg s$^{-1}$, which value was slightly larger than that in the run with the original resolution (\S\ref{sec: remove_de_halpha}).
As the resolution of the smoothed H$\alpha$ image is about 130 pc and not sufficient enough to resolve the HII regions, 
the LF is skewed due to the blending of sources (dashed line in figure \ref{fig: hii_lf}).
Consequently, emission removal from the 24-$\mu$m image may lead to underestimating (overestimating) the relative amount of interarm (arm) emission.
Noting that it is difficult to compensate for this error, we continue the procedure of diffuse emission removal here. In the subsequent analysis of SFR and SFE (\S\ref{sec: sfe}), 
to perform an independent check, SFR is also derived using the H$\alpha$ image alone, besides to the SFR derived by the combination of the H$\alpha$ and 24-$\mu$m.
\par
\begin {figure} []
  \begin {center}
  	\FigureFile(88mm, 100mm){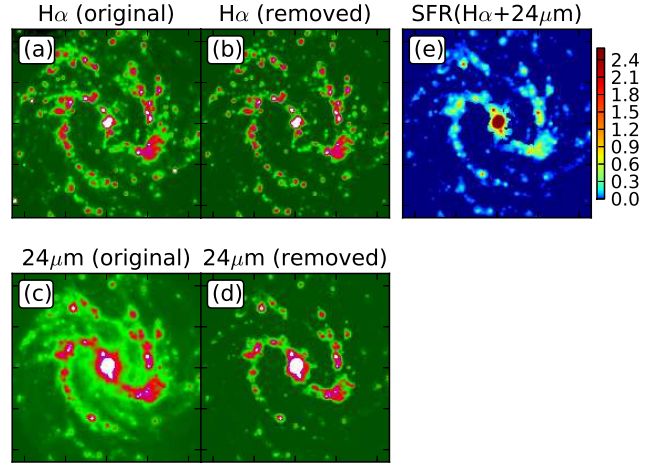}
	\caption {(a) Pseudo-color representation of the original H$\alpha$ image.  (b) Same as (a), but for the background-subtracted H$\alpha$ image. Both (a) and (b) share the same color scaling. (c) Same as (a), but for the original 24-$\mu$m image. (d) Same as (a), but for the background-subtracted 24-$\mu$m image. Both (c) and (d) share the same color scaling.  (e) Image of SFR derived from the linear combinations of the background-subtracted H$\alpha$ and 24$\mu$m images based on the recipe of \cite{Calzetti2007}.}
  \label {fig: remove_de}
  \end {center}
\end {figure}

Using the reference boundaries, the background emission in the 24-$\mu$m image was estimated. 
This estimation was made by replacing each island of pixels identified as discrete sources with the local background emission,
which was determined by two-dimensional surface fitting to the boundary pixels.
After the background image was generated, it was smoothed with 
a median filter with a kernel size of 18$^{''}$ to remove discontinuous edges and any other small-scale features.
Finally, a background-subtracted 24-$\mu$m image was obtained by 
subtracting the smoothed background image from the original image.
Figures \ref{fig: remove_de}(a) and \ref{fig: remove_de}(b) show the original and background-removed 24-$\mu$m images, respectively, with the same color scales.

\subsection{Calculation of Star Formation Rate}
\label{sec: calc_sfr}
The SFR is derived using the the background-subtracted H$\alpha$ and 24-$\mu$m images using the the SFR calibration proposed by \citet{Calzetti2007}.
With the SFR calibration of \citet{Calzetti2007},
the extinction-corrected H$\alpha$ luminosity ($L(\mathrm{H}\alpha_{\mathrm{corr}})$) is expressed as a linear combination of the H$\alpha$ and 24-$\mu$m luminosities:
\begin {equation}
L(\mathrm{H}\alpha_{\mathrm{corr}}) = L\left(\mathrm{H}\alpha\right)+ 
                                         \left(0.031\pm0.006\right)L\left(24{\mu}m\right),
\label{eqn: halpha_24um}
\end {equation}
where $L(\mathrm{H}\alpha)$ and $L$(24$\mu$m) are H$\alpha$ and 24-$\mu$m luminosities, respectively.
Consequently, the SFR is obtained as 
\begin {equation}
\mathrm{SFR} \left(M_{\odot}\mathrm{yr}^{-1}\right) =
5.3\times10^{-42} L(\mathrm{H}\alpha_{\mathrm{corr}})\left(\mathrm{ergs}\ \mathrm{s}^{-1}\right).
\label{eqn: halpha_SFR}
\end {equation}
Figure \ref{fig: remove_de}(e) shows the map of SFR derived with the above equations, using the background-removed H$\alpha$ and 24-$\mu$m images.
\subsection{Radial Distribution of Rate and Efficiency of Star Formation}
\label{sec: radial}
\begin {figure} []
  \begin {center}
  	\FigureFile(88mm, 80mm){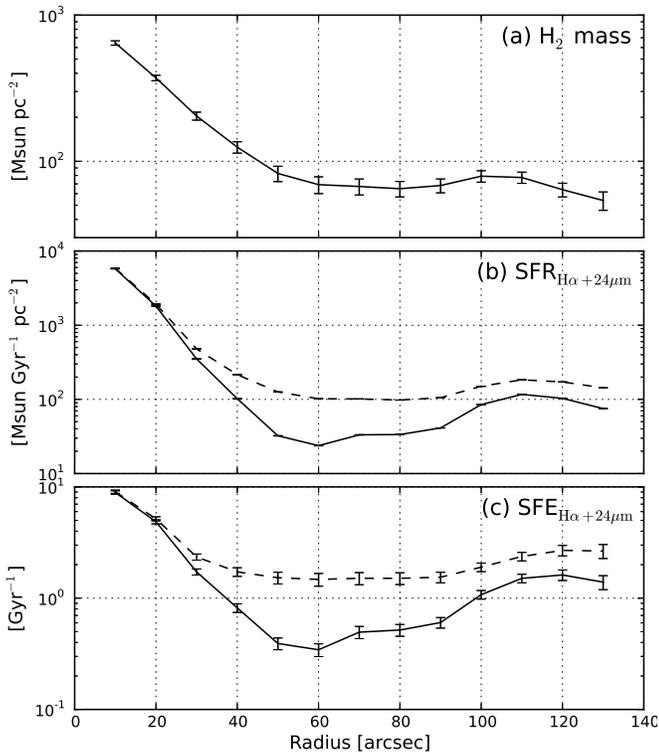}
	\caption {(a) Radial distribution of H$_2$ mass.
    (b) Same as (a), but for SFR derived from the linear combination of H$\alpha$ and 24-$\mu$m images. Solid and dashed lines indicate the SFR with and without background subtraction, respectively. (c) Same as (b), but for SFE ($\equiv$SFR /$M_{\mathrm{H}_{2}}$).
    }
  \label {fig: radial}
  \end {center}
\end {figure}

Figures \ref{fig: radial}(a), (b), and (c) show the radial distributions of H$_2$ mass, SFR, and SFE ($\equiv$SFR /$M$(H$_2$))
, respectively. 
The $^{12}$CO (1--0) integrated intensity is converted into the equivalent H$_2$ mass using the conversion factor of 
$X_{\rm{CO}}$ = 2 $\times$ 10$^{20}$ cm$^{-2}$ (K km s$^{-1}$)$^{-1}$ (\cite{StrongMattox1996}, \cite{Dame2001}).
The SFR is derived with the above equations, both for the background-subtracted and unsubtracted images.
Both the radial distributions of H$_2$ and SFR exhibit two peaks at the center and at the end of the bar. 
The radial distribution of SFE also shows a dip when the diffuse background is removed.

\section{Verifications of the Observational Signs of the Stationary Density Waves}
As summarized in \S\ref{sec: introduction}, 
the extent to which the conventional density wave theory and galactic shock model explain the actual aspects of galaxies is still under active debate;
studies to test two observable influences of the density wave/galactic shock model, namely, the "age gradient" and enhancement of star formation  have often provided compelling results  (e.g., \cite{CepaBeckman1990HaCoRatio, Egusa2004, Tamburro2008, Egusa2009, Foyle2010, Foyle2011, SilvaVilla2012, Louie2013M51Offset}).
Besides these compelling observational results, 
the increased understanding of the non-stationary nature of galactic structures gained via recent simulations (e.g., \cite{Wada2011Interplay, Baba2013}) also reinforces the need for a detailed examination of the two observable influences of galactic structures.
\subsection{Spatial Offsets Between Molecular Clouds and Star Forming Regions}
\label {sec: angoffset}

\begin {figure} []
    \FigureFile(88mm,90mm){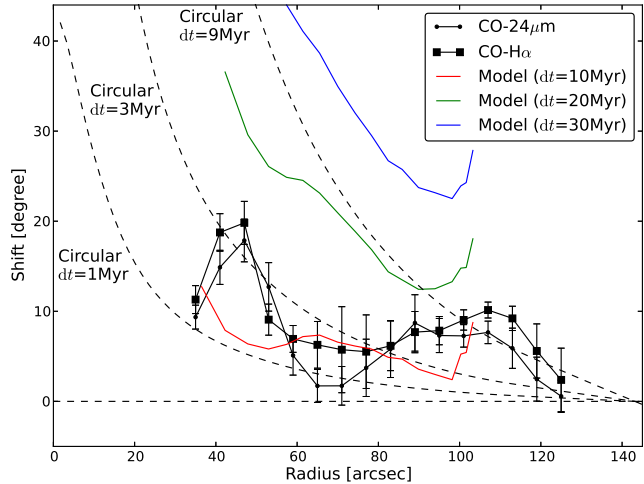}
    \caption {Plot of azimuthal offsets between CO and star formation tracers as a function of the galactocentric radius. Offsets between CO and 24-$\mu$m (dots) and between CO and H$\alpha$ (boxes) are indicated. 
    Dotted, dashed, and dot-dashed lines indicate cases with circular rotation and t$_{delay}$ = 1, 3, and 9 Myr, respectively.
    }
    \label{fig: cc_offset}
\end {figure}

\begin {figure} []
    \FigureFile(72mm,90mm){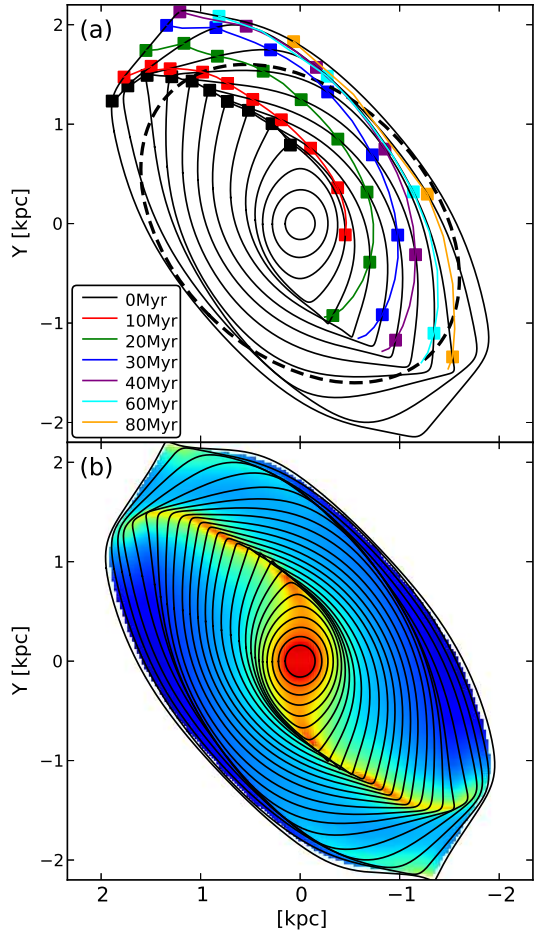}
    \caption {(a) Gaseous orbits in the rotating frame calculated from the potential of M83 (Appendix \S\ref{sec: hydro}). The gas flows in the clockwise direction. The position angle of the bar is set to 225$^{\circ}$.  The dashed ellipse has a semi-major radius of 1.9 kpc, and it indicates the rough extent of the stellar bar.
    Assuming that apocenters of the orbits correspond to the molecular ridge, we traced the movement of material traveling along each orbit in time steps of 10, 20, 30, and 40 Myr.
    (b) Expected density map of material generated by assuming a continuity equation. 
    The solid lines indicate the orbits.
    }
    \label{fig: dorbit}
\end {figure}

As described in the previous section (\S\ref{sec: remove_de_halpha}), bright HII regions that are most probably powered by clustered high-mass star formation are concentrated around the bar and the spiral arms of M83. 
These bright HII regions are also preferentially located downstream of the bar and the arms, at least within the observed region. 

To examine the spatial offsets more quantitatively, the offsets between the CO (1--0) and star formation tracers are analyzed using the angular cross-correlation method \citep{Sheth2002, Tamburro2008, Foyle2011}.
The CO and the diffuse-background-removed SFR tracer (H$\alpha$ and 24-$\mu$m) images are convolved to the common resolution of 12$^{''}$.27, 
stretched to correct for the inclination of the galaxy, and subsequently transformed into the radius-angle plane. 
The pixel size of the radius-angle image is 6$^{''}$ and 1$^{\circ}$ along the radial and azimuthal directions, respectively. 
Each projected SFR tracer image is shifted along the azimuthal direction within a range between -30$^{\circ}$ and 30$^{\circ}$,
and the cross-correlation coefficients between the CO and the shifted SFR images is calculated for each radius.
The peak of the azimuthal profile of the cross-correlation coefficients is fitted with a polynomial function to derive the 
angular shift that maximizes the cross-correlation coefficients.
The angular shift at each radius is taken as an angular offset between the CO and SFR tracers.
The error in the angular shift is estimated by adding noise to each image and running the procedure iteratively.
\par
Figure \ref{fig: cc_offset} shows the radial dependence of the angular offsets between the CO and star formation tracers.
The angular offsets are all positive within the observed region, 
thereby confirming that both the SFR tracers are preferentially located downstream of the CO emission.
The pattern speed determined in \S\ref{sec: patternspeed} locates the CR of M83 at the galactocentric radius of 
141 $\pm$ 14$^{''}$. 
As the observed region is within the CR, the observed positive CO-SF offsets are reasonable, provided that both the bar and arm share the same pattern speed.\par
The angular offsets are often analyzed with the assumption of circular rotation (e.g., \cite{Egusa2004, Tamburro2008, Louie2013M51Offset}). In this case, the angular offset ($\phi(R)$) is simply expressed as 
\begin {equation}
\phi(R) = \left(V_{\mathrm{rot}}(R) / R - \Omega_{\mathrm{p}}\right) \times t_{\mathrm{delay}},
\end {equation} where $V_{\mathrm{rot}}$ denotes rotational velocity,
$\Omega_{\mathrm{p}}$ the pattern speed of the galactic structures,
and $t_{\mathrm{delay}}$ 
the time required for the gas material to travel between the offsets (\cite{Egusa2004}, \cite{Tamburro2008}).
Another interpretation of $t_{\mathrm{delay}}$ is that it represents the time scale required for star-forming regions to emerge from natal molecular clouds.
Previous studies on spiral galaxies (e.g., \cite{Egusa2009, Louie2013M51Offset}) have reported that CO-H$\alpha$ offsets in certain spiral arms are roughly in agreement with the value calculated using the above expression and that $t_{\mathrm{delay}}$ is around 10 Myr, which is comparable to the free-fall time scale of large cloud complexes. 
Three curves with $t_{\mathrm{delay}}$ = 1, 3, and 9 Myr and with $\Omega_{\mathrm{p}} = 57$ km s$^{-1}$ kpc$^{-1}$ are shown in figure \ref{fig: cc_offset}. 
Under the assumption of circular rotation , $t_{\mathrm{delay}}$ needs to to be in the range of 1 to 3 Myr within the bar ($\le85^{''}$). 
However, as this is well below the typical lifetime of the HII regions ($\sim$10 Myr), this value is probably invalid.
\par
For the purpose of modeling the observed angular offsets, we generated a cloud orbit model based on the analytical solution of the gaseous orbits \citep{Wada1994},
which introduces a damping force term (-2$\lambda\dot{R}$) to the equation of motion under weak gravitational perturbation \citep{BinneyTremaine1987}.
We adopted the solution for $m$ = 2 perturbation presented by \citet{Sakamoto1999}, and expanded it to also include $m$ = 4, 6, and 8 components to 
use the gravitational potential distribution estimated from the 2MASS $K_s$-band image (Appendix \S\ref{sec: hydro}).\par
Figure \ref{fig: dorbit}(a) shows the derived orbits. 
The elliptical orbits inside the stellar bar are inclined to each other by small angles and 
form slightly curved offset ridges at the leading sides by orbit crowding.
Outside of the stellar bar, box-shaped ($m$ = 4) orbits become dominant, and the angular shifts in the orbits increase, 
thereby resulting in arm-like structures. 
For the purpose of preference, figure \ref{fig: dorbit}(b) shows the expected density map of 
gas material generated by using the continuity equation. \par
Assuming that the apocenters of the orbits correspond to the molecular ridge, we traced the movement of material that travels along each orbit (boxes on figure \ref{fig: dorbit}a).
Calculating the azimuthal angular offsets from the defined ridge, we compared these with the observed angular offsets shown in figure \ref{fig: cc_offset}.
Although we do not expect the model to completely align with the observed data because the model assumes weak perturbation, 
the model curve with a delay time of 10 Myr is found to be largely in agreement with the observed CO-H$\alpha$, CO-24$\mu$m curves. 
\par
It is notable that the delay time determined here is close to values determined for other spiral galaxies \citep{Egusa2009}. Following the interpretation made in spiral arms, we speculate that most HII regions starts to form inside the peaks of molecular bar, then they migrate along the cloud orbits and after about 10Myr they reach maximum in luminosity. \par
The ``damped'' orbit model 
\citep{Wada1994} also naturally explain the existence of the HII regions located away from the offset ridges toward the leading sides seen in M83 (figure \ref{fig: m83_hiiregions}), as in other barred galaxies 
(e.g., \cite{Sheth2000NGC5383, Sheth2002, Koda2006NGC4303, Zurita2008}). 
\citet{Sheth2000NGC5383} hypothesized that HII regions on the leading sides born from the self-gravitating gas clouds following ballistic orbits. \citet{Koda2006NGC4303} modeled the gas cloud orbits in the barred galaxy NGC 4303 and reported that orbits extending from the ends of the bar can account for the existence of such leading-side HII regions. \par
\par
\begin {figure} []
    \FigureFile(88mm,90mm){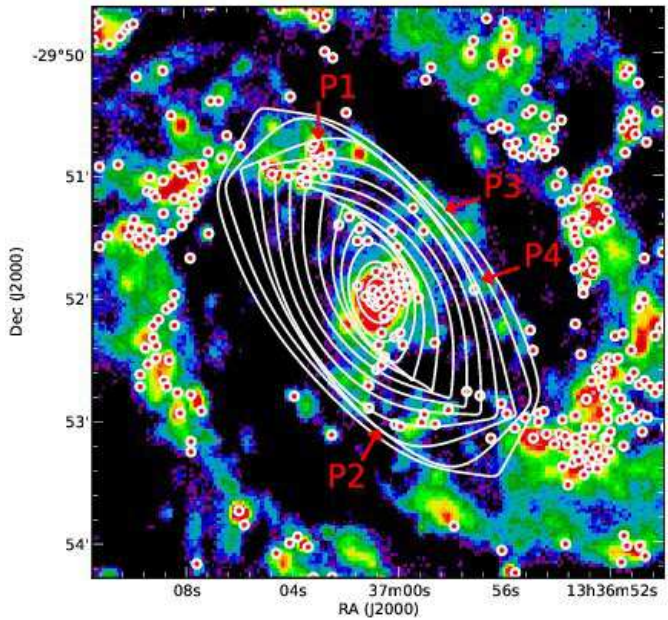}
    \caption {
        {\it GALEX} near-UV (NUV) map of M83 overlaid with the gas orbits in the rotating frame. The circles indicate the bright HII regions with uncorrected H$\alpha$ luminosity over 10$^{37.6}$ erg s$^{-1}$. The solid lines indicate the cloud orbits.
        We note that the outer orbits on the northern side exit from the northern end of the bar and trace the arc of NUV emission which extends in clockwise direction, reaching the northwestern side (from P1 to P3 and P4).
    }
    \label{fig: galex_nuv_dorbit}
\end {figure}

We speculate that this is a case similar to in NGC 4303.
Figure \ref{fig: galex_nuv_dorbit} shows the near-UV map obtained via the {\it GALEX} satellite observation in comparison with the cloud orbits.
The locations of the bright HII regions presented in figure \ref{fig: m83_hiiregions} are also indicated.
The figure clearly indicates that the trajectories of the outer orbits that extend from the ends of the bar pass through the HII regions located at the leading side (P1 and P2).
For the HII regions that were born around the ends of the bar and migrated along the cloud orbits, the required time for travel is 10-30 Myr (figure \ref{fig: dorbit}), and this range does not contradict the typical lifetime of HII regions.
Moreover, star-forming (SF) regions around P3 and P4, which are visible in the UV image but cannot be distinguished in the H$\alpha$ image,
also lie on the same trajectories.
Considering the lifetime of the UV and H$\alpha$ emissions, the age of these SF regions (P3 and P4) could lie between several of tens of Myr to 100 Myr.
Again, it is reasonable to consider that the SF regions travel along the cloud orbits beginning from the ends of the bar since the required time for migration is around 100 Myr (figure \ref{fig: dorbit}).\par
By analyzing the CO-SF offsets and comparing these with the cloud orbits, we observed that 
the distribution of molecular gas and SF regions are in agreement with the conjecture that molecular ridges are the primal sites for star formation. 
Although a certain fraction of star formation can also occur in the interarm regions as they are not completely devoid of molecular gas, 
this mode of star formation may not be efficient.

\par

\subsection{Arm/bar to Interarm Comparison of Star Formation Efficiency}
\label {sec: sfe}
\begin {figure*} []
  \begin {center}
  	\FigureFile(180mm, 100mm){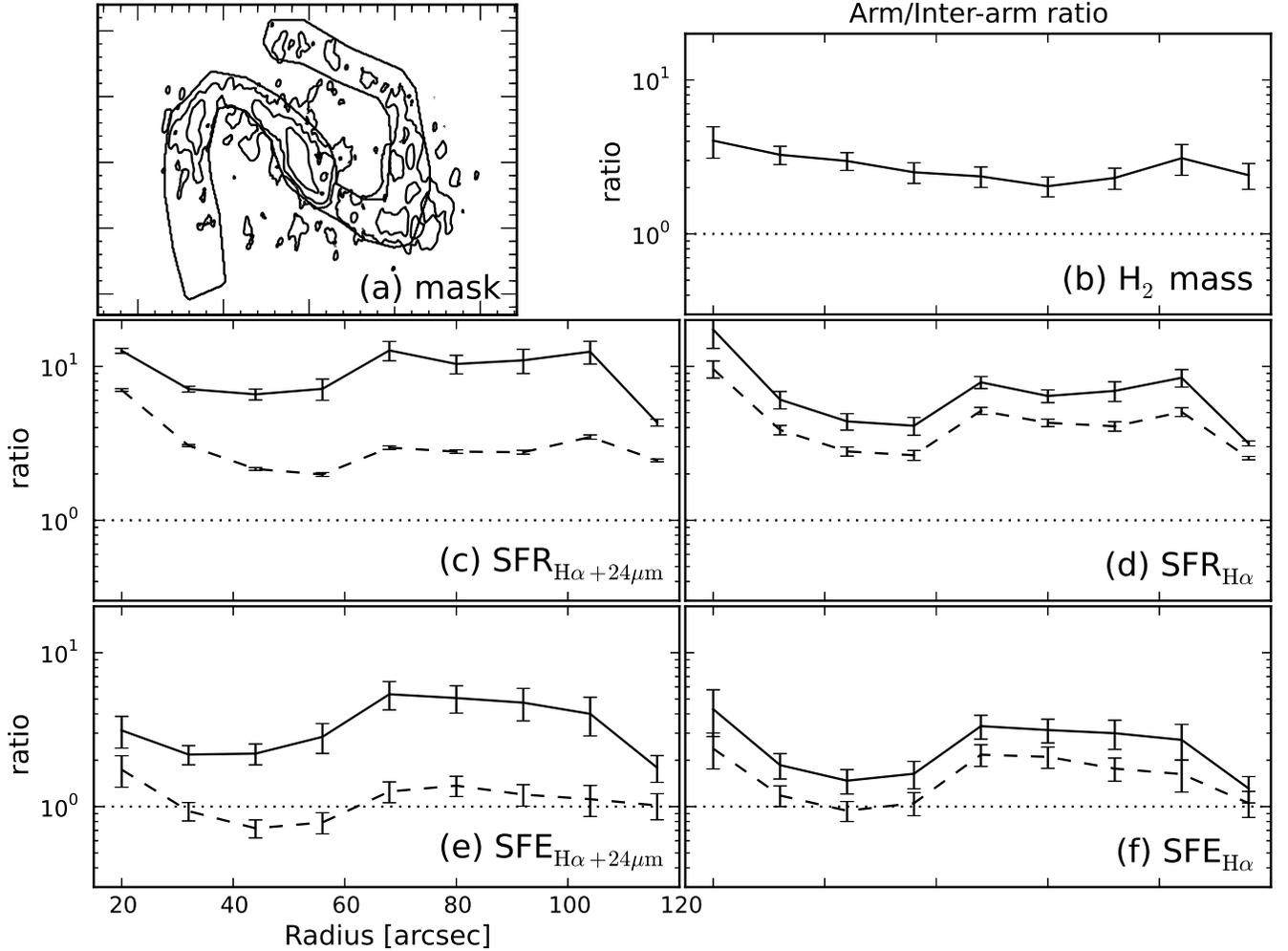}
	\caption{
        (a) Regional mask that separates the arm/bar and interarm regions.
        (b) Radial dependence of the arm-to-interarm ratio of H$_{2}$ mass. (c) Same as (b), but for SFR derived from the combination of H$\alpha$ and 24-$\mu$m images. Solid and dashed lines indicate the case with and without background removal, respectively. (d) Same as (c), but for SFE ($\equiv$SFR/$H_2$ mass). (d) Same as (b) but for the SFR derived from the H$\alpha$ image without correction for internal extinction. (f) same as (d), but for SFE.
    }
  \label {fig: arm_iarm}
  \end {center}
\end {figure*}
Using the SFR derived in \S\ref{sec: sfr}, we now derive the arm-to-interarm contrast of the SFE and test whether 
arm/bar actually enhances the SFE. or only enables concentration of the gas material.
For this purpose, a regional mask that discriminates between the arm+bar and interarm regions (figure \ref{fig: arm_iarm}a) is utilized.
The CO and SFR images are convolved to the common resolution of 12$^{''}$.27, and 
using the mask, we derived the radial dependence of the arm-to-interarm ratio of H$_2$ mass, SFR, and SFE ($\equiv$SFR /H$_2$ mass).
To examine the influence of the diffuse background, SFR and SFE are derived for both image sets with and without background subtraction.
\par
Figures \ref{fig: arm_iarm}(b), \ref{fig: arm_iarm}(c), and \ref{fig: arm_iarm}(e) show the radial dependence of the arm-to-interarm contrast 
for H$_{2}$ mass, SFR, and SFE, respectively.
While arm to interarm contrast of the H$_{2}$ surface mass density ranges from 2 to 3,
that of the SFR with background-removal in the range of 7 to 12, thereby showing higher contrast.
Thus, the contrasts of the SFE are greater than unity (2--5), and this implies that the "triggering" of star formation is ongoing, as predicted by the classical galactic shock models.
We note that, on the other hand, when the background emission is not removed, the SFE contrasts are nearly constant around unity, as previously reported by \citep{Foyle2010} for other grand design spiral galaxies.
\par
As mentioned in \S\ref{sec: remove_de_24um}, because of the coarse resolution of the 24-$\mu$m image ($\sim$130 pc), 
the background removal procedure can lead to the risk of underestimating (overestimating) the relative amount of interarm (arm) emission.
As this effect can erroneously increase the arm-to-interarm ratio of SFE, 
the SFE contrasts derived here are probably an upper limit.
To verify the conclusion that the SFE contrasts are greater than unity even when evaluated with an independent approach, SFR was also derived with H$\alpha$ image alone.
To avoid introducing any other uncertainties, no correction for the internal extinction was made.  Because of the higher amount of gas in the arm/bar compared with that in the interarm regions, the HII regions in the arm/bar should suffer from a higher amount of extinction.
Thus, the extinction-uncorrected value of the SFE contrasts provides the lower limit.\par
Using the background-removed H$\alpha$ image, 
the SFR is calculated using the equation (\ref{eqn: halpha_SFR}).
Figures \ref{fig: arm_iarm}(d) and \ref{fig: arm_iarm}(f) show the arm-to-interarm ratio of the SFR and SFE calculated with the H$\alpha$ image.
The contrasts of the SFE for the background-removed case lie in the range of 1.5-3, and this again supports the enhancement of SFE in the arm/bar.
Although the SFE contrasts derived with the H$\alpha$ image are lower by about a factor of two when compared with those for the H$\alpha$+24$\mu$m image, this difference is not surprising since the arm-to-interarm ratio of the H$_2$ mass, which may be roughly proportional to the extinction amount, exhibits a similar value.
\par
Certain other previous studies on M83 lend further support to our findings.
\citet{Lundgren2008} have generated an SFE map of M83 by estimating the SFR from a combination of H$\alpha$, FUV, and B-band maps and 
calculating the total gas mass from the CO and HI maps. 
In their study, the amount of internal extinction was estimated from the 25$^{''}$-resolution gas mass map.
The SFE map generated by \citet{Lundgren2008} clearly shows an enhancement of SFE in the arm/bar by a factors of 3--5,
which range is in agreement with the results of this study.
\citet{SilvaVilla2012} have derived the SFR based on the resolved stellar populations,
and their results have indicated that the SFR in the arm is about 0.6 dex higher than that of 
the interarm regions. This value is roughly in agreement with the contrasts of the SFR determined here (Figure \ref{fig: arm_iarm}b). \par
Thus, we conclude that the bar/arm in the inner disk of M83 does enhance the SFE, traced by high mass star formation.
The removal of the diffuse light is found to be critical for 
estimating the SFE enhancement since without the removal, 
the arm-to-interarm ratio of the SFE is almost unity, thereby leading to completely opposite conclusion.
Recent studies on the Schmidt--Kennicutt (SK) law also indicate a similar conjecture; 
removing the diffuse background from the SFR tracer images leads to a significant increase in the index of the SK-law \citep{Rahman2011NGC4254, Liu2011KS, Momose2013}. 
 Therefore, the discrepancies between the early studies that report SFE enhancement via H$\alpha$ images \citep{CepaBeckman1990HaCoRatio, LordYoung1990M51, TacconiYoung1990, Knapen1996M10045m} and recent studies that report the absence of SFE enhancement using 24$\mu$m imaging \citep{Foyle2010} may be attributed to the issue of the background removal.

\section{Summary}
The $^{12}$CO (1--0) map of the nearby galaxy M83 generated by observation with the Nobeyama Millimeter Array (NMA) is presented. 
To correct for the lack of the NMA's sensitivity to diffuse emission, the interferometric data are combined with the data obtained using the Nobeyama 45-m telescope. 
The target field of the NMA observations consists of 46 pointings, and it covers the entire extent of the molecular bar and also sections of the spiral arms.
By exploiting the resultant higher angular resolution and sensitivity to extended CO emission, 
the influence of the galactic structures on the molecular gas content is investigated 
in terms of the gas kinematics and the relation between molecular gas and star formation.

\begin {itemize}
    \item{The total-flux-recovered map has a spatial resolution of $\sim$110 pc $\times$ 260 pc, and it clearly resolves galactic structures including the bar and the spiral arms.
    Several interarm clouds with counterparts in the optical images as dust lanes are also detected.
    }
    \item{The position-velocity diagram along the major axis is strongly influenced by non-circular motion induced by the bar. 
    By exploiting the fact that gas streams along the offset ridges and the bar are seen side-on, the pattern speed of the bar ($\Omega_{\mathrm{p}}$) is determined to be 57.4 $\pm$ 2.8 km s$^{-1}$ kpc$^{-1}$. 
    }
    \item{To check the validity of $\Omega_{\mathrm{p}}$,
a simple hydrodynamical simulation is performed to examine the hydrodynamical response of the gas disk. In the simulation, the stellar mass distribution is fixed, and it is calculated from the $K_s$-band image. By comparing the hydrodynamical model predictions with the observed CO data in terms of the kinematics, the adopted $\Omega_{\mathrm{p}}$ value is found to be in agreement with the observed data.
    }
    \item{The star formation rate (SFR) is derived from the linear combination of the H$\alpha$ and the 24-$\mu$m images. 
    Diffuse background emission not related to recent massive star formation is estimated and removed from each image.
    For the H$\alpha$ image, HII regions are identified and removed and subsequently then, background emission is estimated from the residual using the HIIphot code.
    For the 24$\mu$-m image, HII region identification is performed using the smoothed H$\alpha$ image that has a resolution identical to that of the 24-$\mu$m image, 
    and subsequently, by using the detected HII region boundaries as a reference mask, the background 24-$\mu$m emission is estimated.
    }
    \item{
    The luminosity function (LF) of the HII regions identified from the H$\alpha$ image follows a power-law distribution, exhibiting a break at L(H$\alpha$) = 10$^{37.6}$ erg s$^{-1}$.  A similar break in the HII LF has been observed for other galaxies (e.g., \cite{KennicuttHodge1986LMC, Lee2011M51}). The HII regions brighter than the threshold are likely powered by associations of O and B stars, and these bright HII regions are found to be located downstream of the molecular bar/arms. 
    }
    \item{
    Azimuthal angular offsets between CO and H$\alpha$ emission and between CO and 24-$\mu$m images are detected using the angular cross-correlation method.
        The SFR tracers are preferentially located downstream of the CO within the observed region ($R \le 120^{''}$) that not only encompasses the bar but also the inner spiral arms. 
        Since the pattern speed derived in the text places CR as being located 141$\pm$14$^{''}$ away from the center, it is not inconsistent.
        By comparing the amount of the offsets with gas cloud orbits 
        calculated from the stellar potential of M83, 
        it is found that the angular offsets inside the bar roughly correspond to time delay of 10 Myr, which value is in agreement with that determined for other spiral galaxies.
    }
    \item{Finally, using the SFR derived from the background-subtracted H$\alpha$ and 24-$\mu$m images, the arm-to-interarm contrast of the SFE as a function of galactocentric radius is calculated. The SFE contrast is greater than unity (2--5) within the observed region, and suggests that the bar/arms most probably enhance the efficiency of star formation. It is also noteworthy that when the SFE is derived without  background subtraction, the SFE contrast is nearly constant around unity. 
    }
\end {itemize}
In summary, we note that despite recent studies emphasizing the 
importance of the non-steady nature of the spiral arms, 
the CO-SF offsets and the SFE enhancement observed within and around the bar in M83 are in agreement with the predictions of the classical galactic shock model.

\begin{appendix}
    \section{Calculation of Gravitational Potential and Hydrodynamical Modeling of M83}
    \label{sec: hydro}
    The $K_s$-band image retrieved from the data archive of the 2MASS Large Galaxy Atlas \citep{Jarrett2003} is used to trace the stellar mass distribution in M83. 
    Foreground stars, background galaxies, and certain discrete features that are most probably contaminated star forming regions, are removed from the image by fitting each source with a Gaussian function.  
    The ``cleaned image'' is deprojected using the position (225$^\circ$) and inclination (24$^\circ$) angles of the galaxy.\par
\begin {figure} [htbp]
      \begin {center}
      \FigureFile(88mm, 100mm){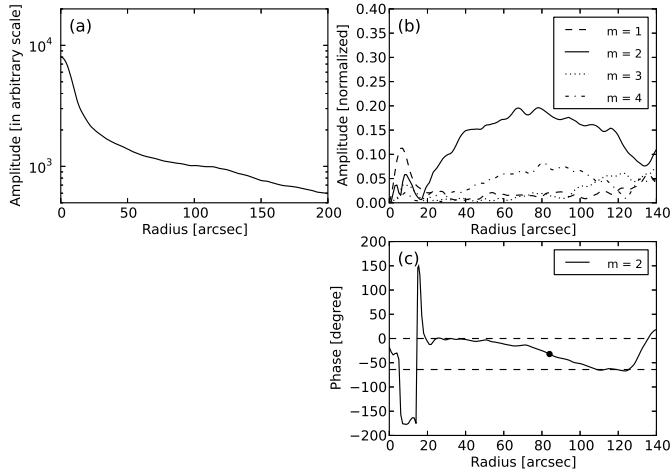}
      \caption {
          (a) Radial distribution of the $m$ = 0 Fourier component of the deprojected $K_s$-band image.  Dashed line indicates the fit with the exponential disk with a scale length of $\sim$2.1 kpc.  (b) Radial distribution of the low order Fourier components ($m$ = 1, 2, 3, and 4). Each Fourier component is normalized by division with the amplitude of $m$ = 0 component.  (c) Phase ($\phi$) of the $m$ = 2 Fourier component as a function of galactocentric radius ($R$).  The phase is nearly constant around zero at $20^{''} \le R \le 50^{''}$ and it gradually changes at the bar-arm transition region to reach $\phi$ = -64$^\circ$ at $R\sim110^{''}$.  The semi-major radius of the bar is determined to be $R$ = 84$^{''}$, where the phase is at the mid-point between $\phi$ = 0$^\circ$ and -64$^\circ$.
  }
      \label {fig: radial_fourier_comps}
      \end {center}
    \end {figure}

    To save the computational costs required for the subsequent hydrodynamical calculation and to smooth out the small scale fluctuations, 
    the deprojected image is Fourier-transformed in polar coordinates and 
    any Fourier components other than low-order even components ($m$ = 0, 2, 4, 6 and 8) are filtered out. 
    Figure \ref{fig: radial_fourier_comps}(b) shows the radial profile of the amplitude of the Fourier components.
    As is clear from the figure,
    the non-axisymmetric perturbations are dominated by even components. 
    Thus, the omission of the odd components does not severely distort the morphology of the galaxy.
    \par
    The stretched and filtered $K_s$-band image is converted into the equivalent stellar surface mass density by 
    applying a constant mass-to-light ratio determined using the maximum disk assumption, which will be referred in the next paragraph.
    To construct a three-dimensional mass distribution,
    a sech-squared law (e.g., \cite{vanderKruit1981}) is assumed for the vertical distribution.
    To determine the scale height, the radial exponential scale length is determined from the radial profile of $m$ = 0 component (figure \ref{fig: radial_fourier_comps}a) and subsequently, a radial-to-vertical scale length ratio of 7 \citep{Kregel2002} is assumed.
    Radial exponential scale length is determined to be $\sim$2.1 kpc and thus, 
    scale height of the sech squared law is determined to be 600 pc, which 
    asymptotically approaches the exponential length of 300 pc over a large distance.\par
\begin {figure} [htbp]
      \begin {center}
      	\FigureFile(80mm, 80mm){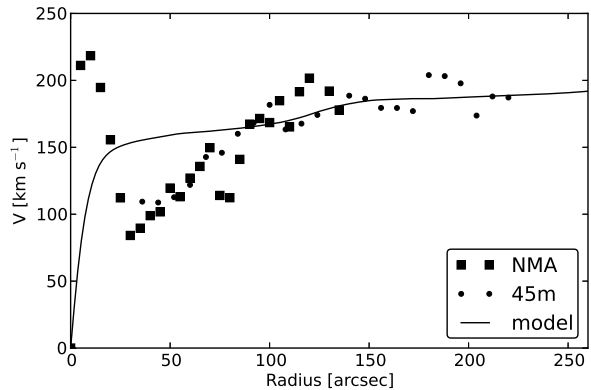}
    	\caption {
Dots and squares represent the observed RC derived from the NMA and the 45m data, respectively.  The solid line indicates the model RC constructed from the stellar distribution estimated from the $K_s$-band image. It is to be noted that within the bar (R $\le 90^{''}$), the "observed" RC is severely affected by non-circular motion.
        }
      \label {fig: rotcurve_compare}
      \end {center}
    \end {figure}

    By solving the Poisson's equation, the distribution of the gravitational potential is calculated.
    Figure \ref{fig: rotcurve_compare} shows the circular rotational velocity calculated from the $m$ = 0 part of the potential distribution.
    The observed RCs derived from the combined NMA+45m data with the envelop tracing method \citep{Sofue1996RC} are also plotted for comparison. 
    The mass-to-light ratio is determined by fitting the observed points of the flat section of the RC ($R \ge 85^{''}$) with the model ($M/L_{K_s} = 0.57 \pm 0.02$ $M_{\odot}/L_{\odot}$).\par

    Using the gravitational potential, a hydrodynamical simulation is performed to reproduce the observed gas kinematics. The Pattern speed of the bar is fixed to $\Omega_{\mathrm{p}}$ = 57 km s$^{-1}$ kpc$^{-1}$.
    A code based on the Advection Upstream Splitting Method (AUSM) scheme is utilized for the calculation, provided by \citet{Nimori2013M83}. 
    The size of the pixel is about 40pc, and no recipe for star formation is included in the calculation.
    The calculation begins with the axisymmetric potential ($m$ = 0) only, and bar/arm perturbations are gradually introduced to avoid introducing artificial numerical noise.
    After this simulation reach the stationary state, the calculation is terminated. 
\end{appendix}
\section*{Acknowledgements}
We are grateful to the NMA and 45-m staff for their help in observations.
We thank Y. Tamura for helping us in the operations of the NMA after 2009, when the common-use operations of the telescope had been ended. This research made use of images provided by the Survey for Ionization in Neutral Gas Galaxies (Meurer et al. 2006) which is partially supported by the National Aeronautics and Space Administration (NASA).
This research has made use of the NASA/IPAC Extragalactic Database (NED) which is operated by the Jet Propulsion Laboratory, California Institute of Technology, under contract with the National Aeronautics and Space Administration. 
This research made use of APLpy, an open-source plotting package for Python hosted at http://aplpy.github.com.

\end {document}